\documentclass[aps,superscriptaddress,twocolumn,floatfix,nofootinbib]{revtex4}

\usepackage{amsmath,amsthm,amsfonts,amssymb,times,bbm,graphicx,color}

\newtheorem{theorem}{Theorem}
\newtheorem{proposition}[theorem]{Proposition}
\newtheorem{lemma}[theorem]{Lemma}
\newtheorem{example}[theorem]{Example}

\newtheorem{conjecture}[theorem]{Conjecture}
\theoremstyle{definition}

\newtheorem{algorithm}[theorem]{Algorithm}

\newcommand{\comment}[1]{}

\newcommand{\newmathcal}[1]{#1}

\def\hr{\mathcal{H}}
\def\C{\mathbb{C}}
\def\ObsDiam{{\rm ObsDiam}}
\def\Sep{{\rm Sep}}

\def\EE{\mathbbm{E}}

\def\R{\mathbb{R}}
\def\idn{\mathbbm{1}}
\def\N{\mathbb{N}}

\DeclareMathOperator{\Tr}{Tr}

\begin{document}

\title{Concentration of measure for quantum states with a fixed expectation value
}

\author{Markus P.\ M\"uller}
\email{mueller@math.tu-berlin.de}
\affiliation{Institute of Mathematics, Technical University of Berlin, 10623 Berlin, Germany}
\affiliation{Institute of Physics and Astronomy, University of Potsdam, 14476 Potsdam, Germany}
\author{David Gross}
\email{david.gross@itp.uni-hannover.de}
\affiliation{Institute for Theoretical Physics, Leibniz University Hannover, 30167 Hannover, Germany}
\author{Jens Eisert}
\email{jense@qipc.org}
\affiliation{Institute of Physics and Astronomy, University of Potsdam, 14476 Potsdam, Germany}
\affiliation{Institute for Advanced Study Berlin, 14193 Berlin, Germany}

\date{October 22, 2010}

\begin{abstract}
Given some observable $H$ on a finite-dimensional quantum system, we
investigate the typical properties of random state vectors
$|\psi\rangle$ that have a fixed expectation value
$\langle\psi|H|\psi\rangle=E$ with respect to $H$. Under some
conditions on the spectrum, we prove that this manifold of quantum
states shows a concentration of measure phenomenon: any continuous
function on this set is almost everywhere close to its mean.  We also
give a method to estimate the corresponding expectation values
analytically, and we prove a formula for the typical reduced density
matrix in the case that $H$ is a sum of local observables. We discuss
the implications of our results as new proof tools in quantum
information theory and to study phenomena in quantum statistical
mechanics. As a by-product, we derive a method to sample the resulting
distribution numerically, which generalizes the well-known Gaussian
method to draw random states from the sphere.
\end{abstract}

\maketitle

\section{Introduction}

The term {\it concentration of measure phenomenon} refers to the
observation that in many high-dimensional spaces ``continuous functions
are almost everywhere close to their mean''. A well-known illustration
is the fact that on a high-dimensional sphere ``most points lie close
to the equator''. In other words, the values of the coordinate functions
concentrate about $0$, their mean. On the sphere, the effect exists
not only for coordinate functions, but for \emph{any} Lipschitz-continuous
function. The result---known as L\'evy's Lemma---has
surprisingly many applications in both mathematics and physics (see
below).

Our main contribution is a ``L\'evy's Lemma''-type concentration of
measure theorem for the set of quantum states with fixed expectation
value.  

More concretely, suppose that
we are given any observable $H=H^\dagger$ with eigenvalues
$\{E_k\}_{k=1}^n$ on $\C^n$.  In the following, we will often call $H$
a ``Hamiltonian'' and $E_k$ the ``energy levels'', but this is not the
only possible physical interpretation.
We fix some arbitrary
value $E$, and we are interested in the set of pure quantum states
with fixed expectation value $E$, i.e.,
\[
   M_E:=\left\{ |\psi\rangle\in\C^n\,\,|\,\, \langle \psi|H|\psi\rangle=E\mbox{ and }\|\psi\|=1\right\}.
\]
Our main Theorem~\ref{MainTheorem1} shows that, subject to conditions on the
spectrum of $H$, any continuous function on $M_E$ concentrates about
its mean.

The motivation for the approach taken here is two-fold. 

\subsection{Motivation 1: The probabilistic method}

Beyond being a geometric
curiosity, the concentration of measure effect  is a crucial
ingredient to an extremely versatile proof technique: the
\emph{probabilistic method} \cite{alon}.
Recall the basic idea. Assume, by way of example, one wants to
ascertain the existence of a state vector $\psi$ on $n$ qubits,
such that $\psi$ is ``highly entangled'' with respect to \emph{any}
bipartition of the $n$ systems into two sets. The problem seems
daunting: There are exponentially many ways of dividing the
composite system into two parts. For any bipartition, we need to make
a statement about the entropy of the eigenvalue distribution of the
reduced density matrix -- a highly non-trivial function. Lastly, in
any natural parametrization of the set of state vectors, a change of
any of the parameters will affect the vast majority of the
constraints simultaneously.

Given these difficulties, it is an amazing fact that the probabilistic
method reduces the problem above to a simple lemma with a schematic
proof (detailed, e.g., in Ref.\ \cite{HaydenLeungWinter,LloydPagels}).
Neither the non-trivial nature of the entropy function, nor the
details of the tensor product space from which the vectors are drawn
enters the proof.  Only extremely coarse information -- the Lipschitz
constant of the entropy and the concentration properties of the unit
sphere -- are needed.

Consequently, proofs based on concentration properties are now
common specifically in quantum information theory.  Examples include
the investigation of ``generic entanglement''
\cite{HaydenLeungWinter}, random coding arguments to assess optimum
rates in quantum communication protocols, state merging
\cite{Merging}, the celebrated counterexample to the additivity
conjecture in quantum information theory \cite{Hastings}, or the
resource character of quantum states for measurement-based computing
\cite{TooMuch1,TooMuch2}. 

The tremendous reduction of complexity afforded by the probabilistic
method motivates our desire to prove measure concentration for other
naturally occurring spaces, besides the sphere. For the set of
``states under a constraint'', Theorem~1 achieves that goal and opens
up the possibility of applying randomized arguments in this setting.

\subsection{Motivation 2: Statistical mechanics}

The second motivation draws from notions of {\it quantum statistical
mechanics}
\cite{Srednicki,Goldstein,Popescu,Reimann,Gogolin,Zanardi,Kollath,Rigol,Exact,Speed}.
The predictions of statistical mechanics are based on ensemble
averages, yet in practice prove to apply already to single instances
of thermodynamical systems. This phenomenon needs to be explained.
It becomes at least plausible if there is a measure concentration
effect on the ensemble under consideration.  Concentration implies
that any observable will give values close to the ensemble mean for
almost every state in the ensemble. This will in particular happen at
almost every point on a sufficiently generic trajectory through the
ensemble. Thus there may be an ``apparent relaxation''
\cite{Popescu,Reimann,Kollath,Rigol,Exact,Speed} even in systems not
in a global equilibrium state.

Recently, several authors realized that it is particularly simple to
state a precise quantitative version of this intuition for ensembles
consisting of random vectors drawn from some subspace
\cite{Popescu,Speed}. However, in the context of statistical
mechanics, it may be more natural to consider sets of states with
prescribed energy expectation value, rather than elements of some
linear subspace.  Indeed, such {\it ``mean energy ensembles''} have
been studied before \cite{BrodyHookHughston, BenderBrodyHook,
FreschMoro, Jiang}. Thus, it is natural to ask whether the
concentration results for linear spaces translate to mean energy
ensembles.

We present both positive and negative results on this problem. Since the
mean energy ensemble and its properties depend on the spectrum of the chosen
observable $H$, so does the degree of measure concentration.
For many spectra typically encountered in large many-body systems, our main theorem yields trivial bounds.
As explained in
Section~\ref{SecNoCon}, this is partly a consequence of the fact that
``L\'evy's Lemma-type'' exponential concentration simply does not exist
for such systems. However, for families of Hamiltonians with, for example,
constant spectral radius, we do get meaningful concentration
inequalities. Therefore, the methods presented in this paper are
expected to have a range of applicability complementary to other
approaches. 

The question whether weakened concentration properties can be proven
for more general many-body systems under energy constraints
remains an interesting problem (see
Section~\ref{SecNoCon}).

\section{Main Results and Overview}
\label{SecNotation}

As stated above, we will analyze the set
\[
   M_E:=\left\{ |\psi\rangle\in\C^n\,\,|\,\, \langle \psi|H|\psi\rangle=E\mbox{ and }\|\psi\|=1\right\}
\]
for some observable $H$ and expectation value $E$.
The set of \emph{all} pure quantum states is a complex sphere in $\C^n$; equivalently, we can view it as the unit sphere $S^{2n-1}$
in $\R^{2n}$. The obvious geometric volume measure on $S^{2n-1}$ corresponds to the unitarily invariant measure on the
pure quantum states~\cite{HaydenLeungWinter}. As we will see below, the set $M_E$ is a submanifold of the sphere
(and thus of $\R^{2n}$); hence
it carries a natural volume measure as well, namely the ``Hausdorff measure''~\cite{Federer} that it inherits
from the surrounding Euclidean space $\R^{2n}$. Normalizing it, we get a natural probability measure on $M_E$.

Our first main theorem can be understood as an analog of L\'evy's Lemma~\cite{Ledoux} for the manifold $M_E$.
It says that the measure on $M_E$ is strongly concentrated, in the sense that the values of Lipschitz-continuous
functions are very close to their mean on almost all points of $M_E$. In some sense, \emph{almost all quantum states
with fixed expectation value behave ``typically''}. To understand the theorem, note that $M_E$ is invariant with
respect to energy shifts of the form $E':=E+s$, $H':=H+s\idn$, such that the new eigenvalues are $E'_k:=E_k+s$.
We then have $M'_{E'}=M_E$, i.e., the manifold of states does not change (only its description does).
We call a function $f:M_E\to\R$ \emph{$\lambda$-Lipschitz} if it satisfies $|f(x)-f(y)|\leq \lambda\|x-y\|$,
where $\|\cdot\|$ denotes the Euclidean norm in $\C^n$.

\begin{theorem}[Concentration of measure]
\label{MainTheorem1}
Let $H=H^\dagger$ be any observable on $\C^n$, with eigenvalues $\{E_k\}_{k=1}^n$, $E_{\text{min}}:=\min_k E_k$,
$E_{\text{max}}:=\max_k E_k$, and arithmetic mean $E_A:=\frac 1 n \sum_k E_k$. Let $E>E_{\text{min}}$ be any value
which is not too close to the arithmetic mean, i.e.,
\[
   E\leq E_A - \frac{\pi(E_{max}-E_{min})}{\sqrt{2(n-1)}}.
\]
Suppose we draw a normalized state vector $|\psi\rangle\in\C^n$ randomly under the constraint
$\langle \psi|H|\psi\rangle=E$,
i.e., $|\psi\rangle\in M_E$ is a random state according to the natural distribution described above.
Then, if $f:M_E\to\R$ is any $\lambda$-Lipschitz function, we have
\begin{equation}
   {\rm Prob}\left\{ |f(\psi)-\bar f|>\lambda t\right\}\leq a \cdot n^{\frac 3 2} e^{-cn \left(t-\frac 1 {4n}\right)^2 +
   2\varepsilon\sqrt{n}},
   \label{eqConMain}
\end{equation}
where $\bar f$ is the median of $f$ on $M_E$, and the constants $a$, $c$ and $\varepsilon$ can be determined in the
following way:
\begin{itemize}
\item Shift the energies by some offset $s$ (as described above) such that $E'_{\text{min}}>0$ and 
\begin{equation}
E'=\left(1+\frac 1 n\right)\left(1+
\frac\varepsilon{\sqrt{n}}\right) E'_H
\label{eqEpsilon}
\end{equation}
with $\varepsilon>0$, where $E'_H$ denotes the harmonic mean energy. The offset may be chosen arbitrarily subject only
to the constraint that the constant $a$ below is positive.
\item Compute $c={3 E'_{\text{min}}}/({32 E'})$ and
\begin{eqnarray*}
	E'_Q&:=&\left(\frac 1 n \sum_k {E'}_k^{-2}\right)^{-\frac 1 2}, \\
	a&=&3040 {E'}_{\text{max}}^2 \left[ {E'}^2 \left(1-\frac {{E'}^2}{\varepsilon^2 {E'}_Q^2}\right)\right]^{-1}.
\end{eqnarray*}	
\end{itemize}
\end{theorem}

The theorem involves an energy offset $s$, shifting all energy levels to $E'_k:=E_k+s$. The idea is to choose this shift
such that $E'\approx E'_H$, i.e.\ such that the energy in question becomes close to the harmonic mean energy (we show in
Lemma~\ref{LemHarmonicMean} below that this is always possible).
Specifically, the theorem demands that $E'$ becomes a bit larger than $E'_H$, resulting in a constant $\varepsilon>0$
defined in eq.~(\ref{eqEpsilon}).
The theorem does not specify $s$ uniquely -- there is some freedom for optimizing over the different possible choices of $s$.
However, there is the constraint that $a>0$, which prevents us from choosing too small values of $\varepsilon$ (indeed,
$a>0$ is equivalent to $\varepsilon>E'/E'_Q$). On the other hand,
$\varepsilon$ should not be too large, because it appears in the exponent in eq.~(\ref{eqConMain}).

To apply the theorem, it is often useful to know the value of the median $\bar f$. Our second main theorem
gives an approximation of $\bar f$ in the limit $n\to\infty$:
\begin{theorem}[Estimation of the median $\bar f$]
\label{MainTheorem2}
With the notation from Theorem~\ref{MainTheorem1}, let $N$ be the full ellipsoid
\[
   N:=\left\{ z\in\C\,\,|\,\, \langle z|H'|z\rangle\leq E'\left(1+\frac 1 {2n}\right)\right\},
\]
and let $f:N\to\R$ be any $\lambda_N$-Lipschitz function. Then, the median $\bar f$ of $f$ on the
energy manifold $M_E$ satisfies
\[
   \left| \bar f -\mathbb{E}_N f \right|\leq\lambda_N\left(\frac 3 {8n} + 15 \left({\frac {E'}{E'_{\text{min}}}\cdot
   \newmathcal{O}\left(n^{-\frac 1 2}\right)}\right)^{\frac12}\right),
\]
where
$\newmathcal{O}\left(n^{-\frac 1 2}\right):=\varepsilon/\sqrt{n} + \ln\left(2a n^{\frac 3 2}\right)/(2 n)$.
\end{theorem}

We proceed by discussing a simple example. Suppose we have a bipartite Hilbert space $A\otimes B$ with
dimensions $|A|=3$ and large, but arbitrary $|B|$, and the Hamiltonian
\begin{equation}
   H=\left(\begin{array}{ccc} 1 & & \\ & 2 & \\ & & 3 \end{array}\right)\otimes \idn_B=:H_A\otimes \idn_B.
   \label{eqH}
\end{equation}
We fix the arbitrary energy value $E=\frac 3 2$, and draw a state $|\psi\rangle\in A\otimes B$ randomly
under the constraint $\langle \psi|H|\psi\rangle=E$, which is $\Tr(\psi^A H_A)=E$. What does Theorem~\ref{MainTheorem1}
tell us about concentration of measure for this manifold of quantum states? To have all positive eigenvalues,
our offset $s$ must be $s>-1$, and the shifted harmonic mean energy becomes 
\begin{equation*}
E'_H(s)=3
\left( \frac 1 {1+s} + \frac 1 {2+s}+\frac 1 {3+s}\right)^{-1}. 
\end{equation*}
The offset $s$ (equivalently, the constant $\varepsilon$) is not specified uniquely by Theorem~\ref{MainTheorem1}; we
try to find a good choice by fixing $\varepsilon$ independently of $n$.
After some trial-and-error, $\varepsilon=2$ turns out to be a good choice
(other values work as well, but not $\varepsilon=1$). The next task is to estimate the shift $s$ which results
from our choice of $\varepsilon=2$; it is determined by the equation 
\begin{equation*}
\frac 3 2+s=\left(1+\frac 1 n\right)\left(1+\frac 2 {\sqrt{n}}\right)E'_H(s), 
\end{equation*}
where $n=3|B|$. It
is difficult to solve this equation directly, but it is easy to see that a solution close to $(-4+\sqrt{7})/3\approx -.45$ exists
for large $n$. This fact helps to gain a rough estimate of $s$ which is sufficient to prove strong concentration
of measure: denote the difference of the left- and right-hand side by $f_n(s)$, then $f_n\left(-\frac 1 2\right)>0$
for all $n\geq 8193$. Since $f_n$ is decreasing, we get $f_n(x)>0$ for all $x\in (-1, -\frac 1 2]$, hence $s>-\frac 1 2$.
The constant 
\begin{equation*}
	c=c(s)=\frac{3(1+s)}{32\left(\frac 3 2 +s\right)} 
\end{equation*}
is increasing in $s$, hence $c\geq c\left(-\frac 1 2\right)
=\frac 3 {64}$, and similarly ${E'_{\text{max}}}/{E'}\leq \frac 5 2$. On the other hand, since $f_n(0)<0$ for all $n\in\N$,
we have $s<0$. Hence we have to consider the expression
${{E'}^2}/({\varepsilon^2 {E'}_Q^2})$ only in the relevant interval $s\in(-\frac 1 2,0)$, where it is decreasing
and thus upper-bounded by $\frac{259}{675}$. Consequently,
$a$ is positive and satisfies $a<30830$. Substituting these expressions into
Theorem~\ref{MainTheorem1}, we get the following result:
\begin{example}
\label{ExMain1}
Drawing random pure state vectors $|\psi\rangle$ under the constraint $\langle\psi|H|\psi\rangle=\frac 3 2$, where
$H$ is the observable defined in eq.\ (\ref{eqH}), we get the concentration of measure result
\[
   {\rm Prob}\left\{ |f(\psi)-\bar f|>\lambda t\right\} \leq 30830\, n^{\frac 3 2}
   e^{-\frac 3 {64} n  \left(t-\frac 1 {4n}\right)^2 +4\sqrt{n}}
\]
for every $\lambda$-Lipschitz function $f$ and all $n\geq 8193$.
\end{example}

It is clear that the amount of measure concentration that we get from Theorem~\ref{MainTheorem1}
depends sensitively on the spectrum of the Hamiltonian $H$. In particular, not all natural
Hamiltonians yield a non-trivial concentration result. For example, in Section~\ref{SecNoCon}, we show
that for a sequence of $m$ non-interacting spins, Theorem~\ref{MainTheorem1} does not give a useful
concentration result in the sense that the corresponding concentration constant $c$ in~(\ref{eqConMain}) will be
very close to zero.
However, we will also prove that this is not a failure of our method, but reflects the fact that there simply
\emph{is no concentration} in that case, at least no concentration which is exponential in the dimension.

In the ``thermodynamic limit'' of large dimensions $n$, the condition on the energy $E$ in Theorem~\ref{MainTheorem1}
becomes $E_{\text{min}}<E<E_A$. From a statistical physics point of view, $E_{\text{min}}$ is the ground state energy of
``temperature zero'', while $E_A$ corresponds to the ``infinite temperature'' energy. Hence, the condition
on $E$ can be interpreted as a ``finite temperature'' condition. However, this condition is no restriction:
if one is interested in concentration of measure for $E_A<E<E_{\text{max}}$, then the simple substitution $H\mapsto -H$
and $E\mapsto -E$ will make Theorem~\ref{MainTheorem1} applicable in this case as well.

In the situation of Example~\ref{ExMain1} above, with Hamiltonian~(\ref{eqH}) on the bipartite Hilbert space
$A\otimes B$ with fixed $|A|$ and large $|B|$, we may ask what the reduced density matrix $\psi^A$ typically looks like.
It is well-known~\cite{HaydenLeungWinter}
that for quantum states \emph{without} constraints, the reduced density matrix is typically close to the maximally
mixed state. To estimate the typical reduced state in our case, we may consider the Lipschitz-continuous functions
$f_{i,j}(\psi):=\left(\psi^A\right)_{i,j}$, that is, the matrix elements of the reduced state. Theorem~\ref{MainTheorem2}
gives a way to estimate these matrix elements by integration over some ellipsoid.

Instead of doing this calculation directly, we give a general theorem below which gives the typical
reduced density matrix in the more general case that the global Hamiltonian $H$ can be written
\begin{equation*}
	H=H_A+H_B, 
\end{equation*}	
i.e., if it describes two systems without interaction.
The Hamiltonian~(\ref{eqH}) corresponds to the special case $H_B=0$. In this case, we get:
\begin{example}
\label{ExMainRed}
Random state vectors $|\psi\rangle$ under the constraint $\langle\psi|H|\psi\rangle=\frac 3 2$, where
$H$ is the observable defined in~(\ref{eqH}), typically have a reduced density matrix $\psi^A$ close to
\[
   \rho_c= \frac 1 {12}Ê\left(
      \begin{array}{ccc}
         5+\sqrt{7} & 0 & 0 \\
         0 & 2(4-\sqrt{7}) & 0 \\
         0 & 0 & -1+\sqrt{7}
      \end{array}
   \right).
\]
More in detail, we have for all $t>0$ and $n\geq 8193$
\begin{eqnarray*}
   {\rm Prob}\left\{\left\| \psi^A-\rho_c\right\|_2> 3\sqrt{8}\left(t+\frac{59}{\sqrt[4]{n}}\right)\right\} \leq 
   369960\, n^{\frac 3 2} \times \\
   \times\enspace e^{-\frac 3 {64} n \left( t-\frac 1 {4n}\right)^2 + 4\sqrt{n}}.
\end{eqnarray*}
\end{example}

This example is a special case of our third main theorem:
\begin{theorem}[Typical reduced density matrix]
\label{MainTheorem3}
Let $H=H_A+H_B$ be an observable in the Hilbert space $A\otimes B := \C^{|A|}\otimes \C^{|B|}$
of dimension $n=|A|\cdot |B|$ with $|A|,|B|\geq 2$. Denote the eigenvalues of $H_A$ and $H_B$
by $\{E_i^A\}_{i=1}^{|A|}$ and $\{E_j^B\}_{j=1}^{|B|}$, and the eigenvalues of $H$
by $E_{kl}:=E_k^A+E_l^B$ respectively. Suppose that the assumptions of Theorem~\ref{MainTheorem1}
hold, and adopt the notation from there, in particular, $E'_k:=E_k+s$ with the energy
offset $s$ specified there.
Then, the reduced density matrix $\psi^A:=\Tr_B|\psi\rangle\langle\psi|$ of random pure state vectors $|\psi\rangle\in A\otimes B$
under the constraint $\langle \psi|H|\psi\rangle=E$ satisfies
\begin{eqnarray*}
   {\rm Prob}\left\{ \left\| \psi^A-\rho_c\right\|_2 > \sqrt{8} |A| (t+\delta)\right\}
   \leq |A|(|A|+1) a n^{\frac 3 2} \times \\
    \times\enspace e^{-cn\left(t-\frac 1 {4n}\right)^2+2\varepsilon\sqrt{n}}
\end{eqnarray*}
for all $t>0$, where the ``canonical'' matrix $\rho_c$ is given by
\[
   \rho_c=\frac{1+\frac 1 {2n}}{n+1}\left(
      \begin{array}{cccc}
         \sum_{k=1}^{|B|} \frac {E'}{E'_{1k}} & 0 & \hdots & 0 \\
         0 & \sum_{k=1}^{|B|} \frac{E'}{E'_{2k}} &  & \vdots \\
         \vdots & & \ddots &  \\
         0 & \hdots & & \sum_{k=1}^{|B|} \frac{E'}{E'_{|A|k}}
      \end{array}
   \right)
\]
and the constant $\delta$ equals
\[
   \delta=\left({\frac{E'}{E'_{\text{min}}}\left(1+\frac 1 n\right)}\right)^\frac12
   \left(
      \frac 3 {8n}+15\left({
         \frac{E'}{E'_{\text{min}}} \newmathcal{O}\left(n^{-\frac 1 2}\right)
      }\right)^\frac12
   \right),
\]
where $\newmathcal{O}\left(n^{-\frac 1 2}\right)=\frac\varepsilon{\sqrt{n}} + \frac{\ln\left(2 a n^{\frac 3 2}\right)}{2n}$.
\end{theorem}

The ``canonical'' matrix $\rho_c$ given above does not immediately have a useful physical interpretation.
In particular, it is \emph{not} in general proportional to $\exp(-\beta H_A)$, i.e., it is not necessarily
the {\it Gibbs state} corresponding to $H_A$ as one might have expected based on intuition
from statistical mechanics (we discuss this point in more detail in Section~\ref{SecNoCon} below).
Note also that $\rho_c$ is not exactly normalized, but it is close to being normalized (i.e.,
$\Tr\rho_c\approx 1$ in large dimensions $n$, which follows from $E'\approx E'_H$).\\

To illustrate the use of Theorem~\ref{MainTheorem3}, we give a \textit{proof of Example~\ref{ExMainRed}}.
We use the notation and intermediate results from the proof of Example~\ref{ExMain1}. The energy shift $s$ depends on the
dimension $n$, i.e., $s=s(n)$, and we have $\lim_{n\to\infty} s(n)=\frac{-4+\sqrt{7}}3$.
More in detail, if we use Mathematica to compute the Taylor expansion of $s(n)$ at $n=\infty$,
we find the inequality
\begin{equation}
   \frac{-4+\sqrt{7}} 3 -\frac{4(35+16\sqrt{7})}{63\sqrt{n}} < s(n) < \frac{-4+\sqrt{7}} 3
   \label{eqIneqShift}
\end{equation}
for all $n\in\N$. The typical reduced density matrix that Theorem~\ref{MainTheorem3} supplies depends
on $n$. It is
\[
   \rho_c^{(n)}:= \frac{\left(1+\frac 1 {2n}\right)\left(\frac 3 2 +s(n)\right)}{n+1}
   \cdot \frac n 3 \left(
      \begin{array}{ccc}
         \frac 1 {1+s(n)} & 0 & 0 \\
         0 & \frac 1 {2+s(n)} & 0 \\
         0 & 0 & \frac 1 {3+s(n)}
      \end{array}
   \right)
\]
which tends to the matrix $\rho_c$ from the statement of Example~\ref{ExMainRed} as $n\to\infty$.
We can use eq.~(\ref{eqIneqShift}) to bound the difference between $\rho_c^{(n)}$ and
$\rho_c$. Using that $\frac{\frac 3 2 +s}{1+s}$ is decreasing in $s$, while $\frac{\frac 3 2 +s}{2+s}$ and
$\frac{\frac 3 2 +s}{3+s}$ are increasing, a standard calculation yields
\begin{equation}
   \left\| \rho_c^{(n)}-\rho_c\right\|_2\leq \frac 4 {\sqrt{n}}
   \label{eqEstimateCanonicals}
\end{equation}
for all $n\geq 829$. Similar calculations can be used to bound the constant $\delta$ from above:
$\delta < \frac{58}{\sqrt[4]{n}}$ for all $n\geq 550$.
Thus, according to Theorem~\ref{MainTheorem3}, it holds
\begin{eqnarray*}
   {\rm Prob}\left\{ \left\|\psi^A-\rho_c^{(n)}\right\|_2 > 3\sqrt{8} \left(t+\frac{58}{\sqrt[4]{n}}\right)\right\}
   \leq 369960 n^{\frac 3 2} \times \\
   \times\enspace e^{-\frac 3 {64} n \left(t-\frac 1 {4n}\right)^2+4\sqrt{n}}
\end{eqnarray*}
for all $n\geq 8193$. The estimate~(\ref{eqEstimateCanonicals}) together with
$3\sqrt{8}\cdot\frac{58}{\sqrt[4]{n}}+\frac 4 {\sqrt{n}} < 3\sqrt{8}\cdot\frac{59}{\sqrt[4]{n}}$ proves the claim
in Example~\ref{ExMainRed}.
\qed

An interesting aspect of Theorem~\ref{MainTheorem3} is that the typical reduced density matrix
does not maximize the entropy locally (if so, it would be the Gibbs state corresponding to $H_A$).
This is expected to have applications in quantum information theory in situations where random bipartite
states with non-maximal entanglement are considered. It will be shown in Section~\ref{SecInvitation}
below that the reduced density matrices maximize a different functional instead which is related
to the determinant.

\section{Implications for Statistical Mechanics}
\label{SecNoCon}
Recently, the concentration of measure phenomenon has attracted a considerable amount of attention
in the context of quantum statistical mechanics. Consider some ensemble of quantum states,
such as the set of all pure quantum states in a certain subspace of the global Hilbert space.
The subspace might be given, for example, by the span of all eigenvectors corresponding to an energy
in some small interval $[E-\Delta E,E+\Delta E]$ with respect to a
given Hamiltonian $H$.

Suppose we are given a single, particular, random state vector $|\psi\rangle$ from
this subspace. What properties will this pure state have? At first one might be tempted to think that there
is very little knowledge available on the properties of the state, and that most properties
should turn out to be random. However, the concentration of measure phenomenon shows that this is
not the case -- almost all the possible state vectors $|\psi\rangle$ will have many properties in common.

Several authors~\cite{Srednicki,Goldstein,Popescu,Reimann,Gogolin}
have recently argued that this property of measure concentration may
help to better understand certain foundational issues of statistical mechanics.
Conceptually, statistical mechanics aims to predict outcomes of
measurements on systems even in the case that we have only very
limited knowledge about the system (say, we only know a few macroscopic
variables). Ensemble averages are employed to make predictions, and
the predictions agree very well with experiment even in \emph{single}
instances of the system. Concentration of measure is then viewed as a
possible theoretical explanation of some aspects of this phenomenon.

As a paradigmatic example, Ref.\ \cite{Popescu} considers the situation of a bipartite
Hilbert space $\hr_S\otimes \hr_E$, consisting of system $S$ and environment $E$. Then the
setting is investigated where the set of physically accessible quantum states is a subspace $\hr_R\subset \hr_S\otimes \hr_E$
(for example, a spectral windows subspace as explained above). If we are given an unknown random
global quantum state vector $|\psi\rangle\in\hr_R$, then what does the state typically look like for the system $S$ alone?
In this case, the postulate of equal apriori probabilities from statistical physics suggests
to use the maximally mixed state on $\hr_R$, that is $\idn/\dim(\hr_R)$, as an ensemble description.
Then, taking the partial trace over the environment will yield a state $\Omega_S=\Tr_E \idn/\dim(\hr_R)$
which may be used by observers in $S$ to predict measurement outcomes.

According to Ref.\ \cite{Popescu}, concentration of measure in the subspace $\hr_R$ proves that \emph{almost
all quantum state vectors $|\psi\rangle\in\hr_R$ have the property that the corresponding reduced state
$\psi^S:=\Tr_E|\psi\rangle\langle\psi|$ is very close to $\Omega_S$.} That is, $\psi^S\approx \Omega_S$
with overwhelming probability. It is then argued in Ref.\ \cite{Popescu} that this result explains why
using the ensemble average $\Omega_S$ is in good agreement with experiment even in the case of
a single instance of the physical system.

While Ref.\ \cite{Popescu} considers a very general situation that does not allow for specifying directly
what the ``typical'' state $\Omega_S$ looks like, Ref.\ \cite{Goldstein} make additional
assumptions that allow to specify $\Omega_S$ in more detail. That is, if the Hamiltonian $H$ is 
\begin{equation*}
	H=H_S+H_E
\end{equation*}	
(that is, there is no or negligible interaction between system and environment), if the
restriction $\hr_R$ is given by a spectral energy window, and if the bath's spectral density
scales exponentially, then $\Omega_S\sim \exp(-\beta H_S)$ for some suitable $\beta>0$.
That is, under these standard assumptions
from statistical mechanics, Ref.\ \cite{Goldstein} argues that the typical reduced state is a {\it Gibbs state}.

These results raise an immediate question: what happens if the restriction is not given by a subspace?
In statistical mechanics, one often considers the situation that an observer ``knows the total energy
of the system'', but does not know the exact microscopic state. In all the papers previously mentioned,
the intuitive notion of ``knowing the energy'' has been translated to the technical statement of
``knowing with certainty that the quantum state is supported in a spectral subspace''.
Obviously, there is at least one natural alternative: knowing the energy might also be read as
``knowing the energy expectation value $\langle\psi|H|\psi\rangle$ of $|\psi\rangle$''.

In fact, this possibility has been proposed by several authors~\cite{BrodyHookHughston, BenderBrodyHook, FreschMoro}
as a possible alternative definition of the ``quantum microcanonical ensemble''
(we call it {\it ``mean energy ensemble''}).
The results in this paper give some information on the applicability of the mean energy ensemble
in statistical mechanics:
\begin{itemize}
\item[1.] As a positive result, we know from Theorem~\ref{MainTheorem1} that the concentration of measure
phenomenon occurs for the mean energy ensemble as well, if that theorem is applicable to the
particular Hamiltonian $H$ and energy value $E$ that is considered. In this case, many interesting
results from Refs.\ \cite{Srednicki,Goldstein,Popescu,Reimann,Gogolin} carry over to the mean energy ensemble.
\item[2.] On the other hand, the ensemble does not seem to reproduce well-known properties of
statistical mechanics, such as the occurrence of the Gibbs state (cf.\ Theorem~\ref{MainTheorem3}).
Hence it seems to describe rather exotic physical situations.
\end{itemize}
There is an intuitive reason why the mean energy ensemble behaves exotically: calculations involving
Theorem~\ref{MainTheorem2} suggest that the $k$-th energy level $|E_k\rangle$ typically has an
occupation proportional to $|\langle\psi|E_k\rangle|^2\sim 1/E'_k$, where $E'_k$ is the corresponding shifted
energy value (cf.\ Theorem~\ref{MainTheorem1}); this is also visible in the form of the typical reduced
density matrix $\rho_c$ in Theorem~\ref{MainTheorem3}.
In particular, typical state vectors $|\psi\rangle$ ``spread out'' a lot
on the small energy levels. This produces a ``Schr\"odinger cat state'' which is in a coherent superposition
of many different energy states. Such states are not normally observed in statistical physics.

However, point 2.\ does not completely rule out the mean energy ensemble as a description
of actual physics, due to the following fact:
\begin{itemize}
\item[3.] Our result is tailor-made for systems with the property that the corresponding
mean energy ensemble concentrates \emph{exponentially in the dimension $n$}, similarly
as in L\'evy's Lemma (corresponding to inequality~(\ref{eqHypo1}) below with $\kappa(n)={\rm const.}$).
However, many natural many-body systems
do not have such a concentration property, which is why Theorem~\ref{MainTheorem1}, \ref{MainTheorem2} and~\ref{MainTheorem3}
do not apply in those cases.
\end{itemize}
As an example, we will now show that the mean energy ensemble does not concentrate exponentially in
the case of $m$ non-interacting spins. Let the energy levels of each spin be $0$ and $+1$.
The total Hilbert space has dimension $n=2^m$. If $k$ is an integer between $0$ and $m$,
then the energy level $k$ is $m\choose k$-fold degenerate. Moreover, suppose that
we are interested in the energy value $E=\alpha m$, where $\alpha\in(0,\frac 1 2)$.
To see if Theorem~\ref{MainTheorem1} is applicable, we determine a rough estimate
of the energy shift $s$ that has to be employed such that $E'\approx E'_H$.
We thus have to find $s>0$ such that
\[
   E'_H=\left( 2^{-m} \sum_{k=0}^m {m\choose k} \frac 1 {k+s}\right)^{-1} \stackrel ! \approx \alpha m +s;
\]
this is only possible if the $(k=0)$-term gives a significant contribution. The conclusion
is that $s$ must be very small, that is, of the order $s\approx \alpha m 2^{-m}$ (and
this conclusion is confirmed by more elaborate large deviations arguments similar to those
discussed below). But then, the constant $c$ in the exponent in~(\ref{eqConMain}) is approximately
\[
   c=\frac{3 E'_{\text{min}}}{32 E'} \approx \frac 3 {32}\cdot \frac{\alpha m 2^{-m}}{\alpha m}=
   \frac 3 {32}\cdot 2^{-m}=\frac 3 {32}\cdot\frac 1 n
\]
such that Theorem~\ref{MainTheorem1} does not give any measure concentration at all.

It turns out, however, that this result is not a failure of our
method, but reflects the fact that there simply \emph{is no
concentration of measure} which is exponential in the dimension $n$ in
this case.  The precise statement below makes use of the \emph{binary
entropy function} 
\begin{equation*}
	H(\gamma) = -\gamma \log_2\gamma - (1-\gamma)\log_2(1-\gamma)
\end{equation*}
defined for $\gamma\in[0,1]$ (c.f.\ \cite{CoverThomas}).

\begin{example}[Non-interacting spins: no exp.\
concentration]\label{ex:noninteracting}
	Suppose we have $m$ non-interacting spins as explained above, and fix
	the energy value $E=\alpha m$ with $\alpha\in(0,\frac 1 2)$. Consider
	a hypothetical concentration of measure inequality
	\begin{eqnarray}
		 {\rm Prob}\{ |f-\bar f|>t\} &\leq& 
		 b\cdot e^{-\kappa(n) t^2} \label{eqHypo1}
	\end{eqnarray}
	for all $1$-Lipschitz functions $f$. Here, $\bar f$ denotes either
	the mean or the median of $f$, $n=2^m$ is the
	dimension of the system and $b>0$ a fixed constant. 
	If such an inequality is to hold, then necessarily 
	\begin{equation}\label{entropyPowerLaw}
		\kappa(n) = o\big(n^{H(\beta)}\big)
	\end{equation}
	for any $\beta> \alpha$. In particular, the optimal exponent in
	(\ref{entropyPowerLaw}) is strictly smaller than one if $\alpha\neq
	\frac12$ and goes to zero as $\alpha\to0$.
\end{example}

\proof
	Let $\gamma$ be such that $\alpha < \gamma < \frac12$. Let
	\begin{equation*}
		L=\sum_{\{i \,|\, E_i < \gamma m\}} \EE[|\psi_i|^2],\qquad
		R=\sum_{\{i \,|\, E_i \geq \gamma m\}} \EE[|\psi_i|^2].
	\end{equation*}
	Normalization gives $L+R=1$ from which we get
	\begin{eqnarray}
		&&\alpha m = E= \sum_i \EE[|\psi_i|^2]\, E_i \geq \gamma m \, R =
		\gamma m (1-L) \nonumber \\
		&\Rightarrow&
		L \geq 1-\frac{\alpha}{\gamma}. \label{Lalphagamma}
	\end{eqnarray}
	We can also bound $L$ using inequality
	(\ref{eqHypo1}). 
	To this end,
	consider the $1$-Lipschitz function $|\psi\rangle\mapsto \Re\psi_i$,
	that is, the real part of the $i$-th component of $|\psi\rangle$ in
	the Hamiltonian's eigenbasis. Due to the invariance of the energy
	manifold with respect to reflections $|\psi\rangle\mapsto
	-|\psi\rangle$, both expectation value and median of this function
	equal zero. If hypothesis~(\ref{eqHypo1}) is true, then it follows
	by squaring that
	\[
		 {\rm Prob}\{(\Re \psi_i)^2>u\}\leq b\cdot e^{-\kappa(n)\,u}
	\]
	and thus
	\begin{eqnarray}
		\EE[(\Re \psi_i)^2 ] 
		&=& \int_{u=0}^\infty u \, \partial_u {\rm Prob}\{(\Re \psi_i)^2\leq u\} \, du
		\nonumber \\
		&=& - \int_{u=0}^\infty u \, \partial_u {\rm Prob}\{(\Re \psi_i)^2> u\}\, du
		\nonumber \\
		&=& \int_{u=0}^\infty {\rm Prob}\{(\Re \psi_i)^2> u\}\, du
		\nonumber \\
		&\leq& \int_{u=0}^\infty b \cdot e^{-\kappa(n) u}\, du =
		\frac{b}{\kappa(n)}, 		\label{byParts}
	\end{eqnarray}
	having used integration by parts. An analogous inequality obviously holds for
	the imaginary part $\EE[(\Im\psi_i)^2]$. From basic information theory
	(e.g., Chapter~11 in Ref.\ \cite{CoverThomas}), we borrow the fact that the number of
	terms  in the definition of $L$ is upper-bounded by
	\begin{equation}\label{typicalSets}
		|\{i\,|\, E_i < \gamma m\}|\leq 2^{m\, H(\gamma)+\log_2 m}.
	\end{equation}
	Combining  (\ref{Lalphagamma}), (\ref{byParts}) and
	(\ref{typicalSets}):
	\begin{eqnarray*}
		&&
		\frac{2 b}{\kappa(n)} 2^{m(H(\gamma)+\frac1m \log_2 m)} \geq L \geq 1 -
		\frac{\alpha}{\gamma} \\
		&\Rightarrow&
		\kappa(n) \leq
		\frac{2 b}{1-\frac{\alpha}{\gamma}} 
		n^{H(\gamma)+\frac1m \log_2m} = o\big(n^{H(\beta)}\big)
	\end{eqnarray*}
	for every $\beta>\gamma$. 
\qed

It is highly plausible that similar results hold for the mean energy
ensemble of many other many-body systems: the best possible rate of
concentration (such as the right-hand side in~(\ref{eqHypo1})) is not
$\exp(-cnt^2)$, but at most $\exp(-c n^p t^2)$ with energy-dependent
exponent $p<1$. Deciding whether this upper bound can be achieved
remains an interesting open problem. Indeed, we end the present
section by sketching a possible route for tackling this question.

The proof of Example~\ref{ex:noninteracting} uses the coordinate
functions $\psi\mapsto\Re\psi_i$ as an example for continuous
functions without strong concentration properties. We conjecture that
this is already the worst case, i.e.\ that no function with
Lipschitz-constant equal to one ``concentrates less'' than
the ``most-spread out'' of the coordinate functions.

A strongly simplified version of this conjecture is easily made
precise. We restrict attention to real spaces and linear functions 
\begin{equation*}
	\psi \mapsto \sum_i c_i \psi_i
\end{equation*}
of the state vectors. Here, the $\psi_i$'s are the coefficients of $\psi$ in
the eigenbasis of the Hamiltonian and the $c_i$'s are arbitrary 
coefficients subject to the normalization constraint $\sum_i c_i^2 =
1$. The latter constraint ensures that the Lipschitz constant of the
linear funtion is one.
If the vectors $\psi$ are drawn from an energy ensemble $M_E$, we have the
elementary estimate
\begin{eqnarray*}
	\operatorname{Var}[\sum_i c_i \psi_i ] 
	&=&\sum_{i,j} c_i c_j \mathbbm{E}[\psi_i \,\psi_j] 
	=\sum_{i} c_i^2 \mathbbm{E}[\psi_i^2] \\
	&\leq&\max_{i}  \mathbbm{E}[\psi_i^2] = 
	\max_i \operatorname{Var}[\psi_i],
\end{eqnarray*}
having made use of the fact that $M_E$ is invariant under the
transformation which takes $\psi_i \mapsto -\psi_i$ for some $i$ while
leaving the other coordinates fixed.

Upper bounds on the variance of a random variable are sufficient to
establish simple concentration estimates (by means of Chebychev's
inequality). From that point of view, we have proven that coordinate
functions show the least concentration among all linear functions,
according to methods based on second moments alone. Generalizing this
observation to more general functions and higher moments would
allow us to restrict attention to the eigenbasis of the Hamiltonian,
thus ``taking the non-commutativity out of the problem''. Conceivably,
this would constitute a relatively tractable path to 
a more complete
understanding of concentration in typical many-body systems.

\section{Invitation: a simple Hamiltonian}
\label{SecInvitation}
As a preparation for the proof in the next section, we give a particularly simple example of a
Hamiltonian which admits a more direct proof of concentration of measure; in the meantime, we
will also see that the ``harmonic mean energy'' from Theorem~\ref{MainTheorem1} appears naturally.
Consider a bipartite quantum system on a Hilbert space $\hr=\hr_A\otimes\hr_B$, with dimensions
$|A|:=\dim\hr_A$ and $|B|:=\dim\hr_B$. We may assume without restriction that $|A|\leq |B|$. We are
interested in the manifold of state vectors $|\psi\rangle\in\hr$ with average energy $\langle\psi|H|\psi\rangle=E$,
where $H=H^\dagger$ is some Hamiltonian on $\hr$.
What happens if we draw a state $|\psi\rangle$ from that submanifold at random? Instead of
studying this question in full generality (which we will do in Section~\ref{SecConcentration}),
we start with the simple example
\[
   H=H_A\otimes \idn,
\]
where $H_A=H_A^\dagger$ is an observable on $\hr_A$ alone. This is a special
case of a more general bipartite Hamiltonian $H=H_A+H_B= H_A\otimes \idn + \idn\otimes H_B$
without interaction as studied in Theorem~\ref{MainTheorem3}. A nice consequence is that
\[
   \langle\psi|H|\psi\rangle=E\Leftrightarrow \Tr(\psi^A H_A)=E,
\]
that is, the constraint depends on the reduced density matrix $\psi^A:=\Tr_B |\psi\rangle\langle\psi|$
alone.

The unitarily invariant measure $\nu$ on the pure quantum states on the global Hilbert space $\hr$
induces in a natural way a measure $\nu_A$ on the density matrices in $\hr_A$: if $S$ is some measurable
subset of the density matrices, then
\[
   \nu_A(S):=\nu\left(\left\{ |\psi\rangle\,\,|\,\, \Tr_B |\psi\rangle\langle\psi|\in S\right\}\right)
   =\nu\left(\Tr_B^{-1}(S)\right).
\]
Formally, the measure $\nu_A$ is the \emph{pushforward measure}~\cite{Gromov} 
of the unitarily invariant measure $\nu$
on the pure states of $\hr_A\otimes \hr_B$ with respect to the map $\Tr_B$; that is,
$\nu_A=\left(\Tr_B\right)_*(\nu)$.

Due to the simple form of the Hamiltonian $H$, we may calculate probabilities with respect to $\nu_A$.
The probability density distribution corresponding to $\nu_A$ is invariant with respect to unitaries
on $\hr_A$, but it depends on the eigenvalues $t=(t_1,\ldots,t_{|A|})$ of the reduced density matrix
$\psi^A$. The relation is~\cite{LloydPagels,ZycPaper}:
\begin{equation}
   d\nu_A(t)=z^{-1}\delta\left(1-\sum_{i=1}^{|A|} t_i\right)\prod_{i=1}^{|A|} t_i^{|B|-|A|} \prod_{i<j\leq |A|} (t_i-t_j)^2 dt,
   \label{EqEVDist}
\end{equation}
where $z$ is the normalization constant and $dt=dt_1\ldots dt_{|A|}$.

We will now study the simplest case $|A|=2$, where
\[
   H_A=\left(\begin{array}{cc} E_1 & 0 \\ 0 & E_2 \end{array}\right),
\]
and we may assume without loss of generality that $E_1>E_2$. We fix some arbitrary energy value
$E$ between $E_1$ and $E_2$. Since the average energy $E_A:=\frac 1 2(E_1+E_2)$
is the energy of an ``infinite temperature'' Gibbs state, we additionally assume that $E< E_A$,
which is no restriction, but saves us some distinction of cases. Thus, $E_2<E< E_A<E_1$.

The state space is now the Bloch ball; that is, the unit ball in $\R^3$. The distribution $d\nu_A$ has been calculated for the case $|A|=2$
already by Hall~\cite{Hall}. In ordinary spherical coordinates, it can be written
\[
   d\nu_A(r)=\rho_{|B|}(r) r^2 dr\, d\varphi\, d\theta,
\]
where $\rho_{|B|}(r)=c_B (1-r^2)^{|B|-2}$, and
$c_B$ is some normalization constant. It can be derived from eq.~(\ref{EqEVDist}) by using that points $r$ in the Bloch
ball correspond to density matrices with eigenvalues $t_1=\frac 1 2(1+\|r\|)$ and $t_2=\frac 1 2(1-\|r\|)$.
If $|B|=2$, this measure becomes the usual Euclidean measure in the Bloch ball,
corresponding to the Hilbert-Schmidt measure~\cite{ZycPaper}. In general,
\[
   \nu_A(S)=\iiint_S \rho_{|B|}(r) r^2 dr\,d\varphi\,d\theta=\iiint_S \rho_{|B|}(|x|) dx.
\]
We write $\rho_B(x):=\rho_{|B|}(|x|)$ for $x\in\R^3$.

It is easy to see (and we will show this below) that the subset of mixed states $\psi^A$ in the Bloch ball with $\Tr(\psi^A H_A)=E$
is the intersection of a plane with the Bloch ball; that is, a disc $K_E$. To determine probabilities,
we need to compute the area $\mu(X)$ of two-dimensional subsets $X\subset K_E$. In light of the density $\rho_B$ introduced above,
it is tempting to use the term $\iint_X \rho_B(x)\, dx$ as the measure of $X$. The normalized version is a probability measure:
\begin{equation}
   {\rm Prob}(X):=\frac{\iint_X \rho_B(x)\, dx}{\iint_{K_E} \rho_B(x)\, dx}.
   \label{eqDefProbAgain}
\end{equation}
It turns out that this measure agrees with the normalized geometric volume measure ${\rm Prob}$ that we use elsewhere in this paper (for example
in Theorem~\ref{MainTheorem1}). We discuss this fact in detail below after the proof of Proposition~\ref{PropInvitation}.
Using this identity and deferring its justification to below, we get a concentration of measure result:
\begin{proposition}
\label{PropInvitation}
The reduced density matrix $\psi^A:=|\psi\rangle\langle\psi|$ concentrates exponentially on the canonical state
\[
   \rho_c:=\left(
      \begin{array}{cc}
         \frac{E-E_2}{E_1-E_2} & 0 \\
         0 & \frac{E_1-E}{E_1-E_2}
      \end{array}
   \right),
\]
that is,
\[
   {\rm Prob}\left\{ \|\psi^A-\rho_c\|_1\geq\varepsilon\right\}
   \leq \exp\left[
      -\varepsilon^2\frac{(|B|-1)(E_1-E_2)^2}{4(E_1-E)(E-E_2)}
   \right]
\]
for all $\varepsilon\geq 0$.
\end{proposition}
\proof
It follows from the condition $\Tr(\psi^A H_A)=E$ that $\psi^A$ must have diagonal elements
\begin{equation*}
	d_1:=\frac{E-E_2}{E_1-E_2},\,\,
	d_2:=\frac{E_1-E}{E_1-E_2}, 
\end{equation*}
where $d_1<d_2$.
By Schur's Theorem~\cite{Bhatia}, the eigenvalues $t_1$ and $t_2$ of $\psi^A$ must
majorize the diagonal elements, that is, $(d_2,d_1)\prec (t_2,t_1)$, hence $t_2=\|\psi^A\|_\infty\geq d_2$.
Thus, among all states $\psi^A$ satisfying the energy condition, the canonical state $\rho_c$
is the ``most mixed'' one, in the sense that its radius in the Bloch ball is the smallest possible.

Geometrically, this has the following interpretation. Translating the condition $\Tr(\psi^A H_A)=E$ to the Bloch ball
representation, that is, representing density matrices $\psi^A$ by vectors $r\in\R^3$, we get
\[
   r\cdot\left(\begin{array}{c} 0 \\ 0 \\ E_1-E_2\end{array}\right)=2(E-E_A).
\]
This defines a plane with 
\[
	r_z=\frac{2(E-E_A)}{E_1-E_2}<0 
\]
in $\R^3$; the intersection
of that plane with the Bloch ball gives the set of density matrices that fulfill the energy
condition. This set is a disc $K_E$, with the vector representation of $\rho_c$ at its center.
Since the trace distance $\|\cdot\|_1$ on the density matrices corresponds to the Euclidean distance
in the Bloch ball, the set of states $\psi^A$ satisfying the energy condition with $\|\psi^A-\rho_c\|_1\geq\varepsilon$
corresponds to an annulus in $K_E$ with inner radius $\varepsilon$; we denote this annulus by $K_E(\varepsilon)$. 
Hence
\[
   {\rm Prob}\left\{ \|\psi^A-\rho_c\|_1\geq\varepsilon\right\}=
   \frac{\iint_{x\in K_E(\varepsilon)} \rho_B(x) dx}
   {\iint_{x\in K_E(0)} \rho_B(x) dx}.
\]
By elementary integration, we get
\begin{eqnarray*}
   \int_{x\in K_E(\varepsilon)} \rho_B(x) dx&=&\int_0^{2\pi} d\varphi\int_\varepsilon^{\sqrt{1-r_z^2}}ds\, \cdot s
   \rho_B\left(
      \begin{array}{c}
         s\cos\varphi\\ s \sin\varphi \\ r_z
      \end{array}
   \right)\\
   &=&2\pi c_B\int_\varepsilon^{\sqrt{1-r_z^2}} s(1-s^2-r_z^2)^{|B|-2} ds\\
   &=&\frac{\pi c_B}{|B|-1}\left(1-r_z^2-\varepsilon^2\right)^{|B|-1}.
\end{eqnarray*}
Since $K_E=K_E(0)$, this yields
\begin{eqnarray*}
   {\rm Prob}\left\{ \|\psi^A-\rho_c\|_1\geq\varepsilon\right\}&=&\left(
      \frac{1-r_z^2-\varepsilon^2}{1-r_z^2}
   \right)^{|B|-1}\\
   &=&\left(1-\frac{\varepsilon^2}{1-r_z^2}\right)^{|B|-1}\\
   &\leq& \exp\left(-(|B|-1)\frac{\varepsilon^2}{1-r_z^2}\right).
\end{eqnarray*}
Substituting 
\[
1-r_z^2=\frac{4(E_1-E)(E-E_2)}{(E_1-E_2)^2} 
\]
proves the claim.
\qed

In this paper, the measure that we are interested in is the Hausdorff volume measure $\mu_{M_E}$ on the energy
manifold $M_E$. To compute the correct probabilities when restricting to the Bloch ball by partial trace,
we need to invoke the pushforward of $\mu_{M_E}$ with respect to $\Tr_B$. So is the probability
measure ${\rm Prob}$ that we have defined above in eq.~(\ref{eqDefProbAgain}) equal to this pushforward measure, i.e., does
\[
   {\rm Prob}\stackrel{?}{=} \left(\Tr_B\right)_*(\mu_{M_E})
\]
hold? Fortunately the answer is ``yes'', but this is not directly obvious. First we observe that ${\rm Prob}$
may be interpreted as an ``energy shell measure'' in the following sense.
The submanifold measure from eq.~(\ref{eqDefProbAgain})
can be given by a limit, in a spirit similar to the definition of the ``Minkowski content'' (cf. Ref.\ \cite{Federer}):
For $X\subset K_E$, denote by $U_\delta^-(X)$
the set of matrices $\psi^A$ that are $\delta$-close to $X$, and have an energy expectation value of $\Tr(\psi^A H_A)< E$.
This is half of the $\delta$-neighborhood of $X$. Then, up to a normalization constant,
\begin{eqnarray*}
   {\rm Prob}(X)&\sim& \iint_X \rho_B(x) dx\\
   &=&\lim_{\delta\to 0} \frac 1 \delta \iiint_{U_\delta^-(X)} \rho_B(x)dx \\
   &=&\lim_{\delta\to 0}\frac 1 \delta \nu_A\left(U_\delta^-(X)\right) \\
   &=& \lim_{\delta\to 0} \frac 1 \delta \nu\left(\Tr_B^{-1}(U_\delta^-(X))\right).
\end{eqnarray*}
But $U_\delta^-(K_E)$ is a ``slice'' of the Bloch ball in between two parallel planes. The set
consists of those matrices $\psi^A$ determined by the inequality $E-\varepsilon<\Tr(\psi^A H_A)<E$,
where $\varepsilon>0$ is some energy difference corresponding to $\delta$. In particular,
\[
   \Tr_B^{-1}\left(U_\delta^-(K_E)\right)=\left\{|\psi\rangle\,\,|\,\, \langle \psi|H|\psi\rangle\in(E-\varepsilon,E)\right\}.
\]
This is an energy shell. Hence it is basically the uniform distribution on this energy shell on the
pure states which, by taking the partial trace and the limit $\varepsilon\to 0$, generates our
probability measure ${\rm Prob}$.

In the general case of arbitrarily many different energy levels $\{E_k\}_{k=1}^n$, this energy shell
has different ``widths'' at different points $x\in M_E$; in the limit $\delta\to 0$, this width is proportional to
$\|P_S \nabla E(x)\|^{-1}$, where $P_S$  denotes the projection onto the sphere's tangent space at $x$.
Hence the corresponding ``energy shell measure''
does not in general equal the geometric (Hausdorff) measure ${\rm Prob}$ used elsewhere in this paper,
which arises from an analogous limit procedure, but starting with the uniform
distribution in an \emph{$\varepsilon$-neighborhood} of $M_E$.

However, here we are in a very special situation: we only have \emph{two different energy levels} $E_1$ and
$E_2$ which are highly degenerate. In this case, it turns out that $\|P_S\nabla E(x)\|$ is constant along $M_E$.
To simplify the argument, we double the dimensions and work in real space $\R^n$; the case $n=2|A|\, |B|$
applies to Proposition~\ref{PropInvitation} above.
\begin{lemma}
Consider the real energy manifold
\[
   M_E:=\left\{ x\in\R^n\,\,\left|\,\, \sum_{k=1}^n E_k x_k^2 = E,\quad \sum_{k=1}^n x_k^2=1\right.\right\}
\]
in the special case that there are \emph{only two different energy levels}, i.e.\ there exist $\mathcal{E}_1,\mathcal{E}_2$
such that $E_1=E_2=\ldots =E_m=\mathcal{E}_1$, and $E_{m+1}=\ldots=E_n=\mathcal{E}_2$. Furthermore, assume that
$\mathcal{E}_1 < E < \mathcal{E}_2$. Let $P_S$ denote the projection onto the tangent space of the unit sphere, and let
$E(x):=\sum_{k=1}^n E_k x_k^2$. Then,
$\|P_S \nabla E(x)\|$ is constant on $M_E$.
\end{lemma}
\proof
Direct calculation yields
\[
   \|P_S \nabla E(x)\|^2 = 4 E^2(x) - 4  E(x)^2,
\]
where $E^2(x):=\sum_{k=1}^n E_k^2 x_k^2$. Since
\[
   E(x)=\left(\sum_{k=1}^m x_k^2\right) \mathcal{E}_1 + \left(1-\sum_{k=1}^m x_k^2\right)\mathcal{E}_2
\]
equals $E$ on all of $M_E$, it follows that $\sum_{k=1}^m x_k^2$ is constant on $M_E$. But then,
\[
   E^2(x)=\left(\sum_{k=1}^m x_k^2\right) \mathcal{E}_1^2 + \left(1-\sum_{k=1}^m x_k^2\right)\mathcal{E}_2^2
\]
is also constant on all of $M_E$.
\qed

Hence, in our case of only two different energy levels, the energy shell has constant width everywhere, such that the measure defined in
eq.~(\ref{eqDefProbAgain}) indeed agrees with the geometric measure that we use elsewhere in the paper.

Proposition~\ref{PropInvitation} shows that a typical reduced density matrix is close to $\rho_c$; that is, it
is diagonal in $H_A$-basis, and it can be written $\rho_c=e^{-\beta H_A}/Z$ with $Z:=\Tr(e^{-\beta H_A})$
and some appropriate ``inverse temperature'' $\beta>0$. Hence it is a Gibbs state.

This result suggests that the local state always concentrates on a Gibbs state when $H=H_A\otimes\idn$,
also in the more general case $|A|=\dim\hr_A\geq 3$. But this guess is false, as we have already
shown in Theorem~\ref{MainTheorem3} and Example~\ref{ExMainRed} -- the local density matrix always
commutes with $H_A$, but it is not in general the corresponding Gibbs state.

In light of the calculation above, it is now easy to give an intuitive explanation for this fact.
In analogy to the previous proposition
for $|A|=2$, we expect that the distribution of eigenvalues as given in eq.~(\ref{EqEVDist}) will dominate
the concentration of measure to some local ``canonical'' state. It also seems reasonable to assume
(and can be verified numerically) that the $(t_i-t_j)^2$-terms do not contribute much for large $|B|$,
and that the other terms exponential in $|B|$ dominate. But these terms are
\[
   \prod_{i=1}^{|A|} t_i^{|B|-|A|}=(\det\psi^A)^{|B|-|A|},
\]
and so we conclude that the reduced density matrix should concentrate on the one $\psi^A$
which maximized the previous expression. This suggests the following conjecture:
\begin{conjecture}
Suppose that $H=H_A\otimes \idn$ is a Hamiltonian on a bipartite Hilbert space $\hr=\hr_A\otimes\hr_B$
with fixed $|A|:=\dim\hr_A$ and varying $|B|:=\dim\hr_B$. Then typical quantum states $|\psi\rangle\in\hr$
with fixed mean energy $\langle\psi|H|\psi\rangle=E$ have the
property that $\psi^A:=\Tr_B|\psi\rangle\langle\psi|$ concentrates exponentially on the determinant maximizer
\[
   \rho_c:=\arg\max \left\{ \det\psi^A\,\,|\,\, \Tr(\psi^A H_A)=E\right\},
\]
given the maximizer is unique.
\end{conjecture}
We will now see that the prediction of this conjecture is consistent with Theorem~\ref{MainTheorem3}.
To this end, we now compute the determinant maximizer $\rho_c$ explicitly.
First of all, it is easy to see that $\rho_c$ must be diagonal if written in the basis
of $H_A$, i.e., $[\rho_c,H_A]=0$: Suppose $\sigma$ is any density matrix with $\Tr(\sigma H_A)=E$,
and let $\{\sigma_{k,k}\}_{k=1}^{|A|}$ denote the diagonal elements. Then $\Tr(\sigma H_A)=\sum_k \sigma_{k,k} E_k =E$,
where $E_k$ denotes the eigenvalues of $H_A$. Now if $\tilde\sigma$ is the matrix with diagonal elements $\sigma_{k,k}$
and other entries zero, then $\Tr(\tilde\sigma H_A)=E$ is still true, but by the Hadamard determinant theorem~\cite{Bhatia}
\[
   \det\sigma \leq \prod_{k=1}^{|A|} \sigma_{k,k} = \det\tilde\sigma.
\]
Hence the maximizer is diagonal; it remains to determine its diagonal elements $(\lambda_1,\ldots,\lambda_{|A|})$.
We have to maximize $\sum_k \ln\lambda_k$ subject to the constraints $\sum_k \lambda_k =1$ and $\sum_k \lambda_k E_k=E$.
It turns out to be difficult to do that directly, so we drop the normalization condition $\sum_k\lambda_k=1$ for the moment
and solve the resulting equation
\[
   \frac{\partial}{\partial \lambda_i} \sum_k \ln\lambda_k - \lambda \frac\partial{\partial \lambda_i} \sum_k \lambda_k E_k =0,
\]
where $\lambda$ is the Lagrange multiplier. This gives $\lambda=\frac {|A|} E$, and $\lambda_i = \frac E {|A|} \cdot \frac 1 {E_i}$.
Since this distribution is not automatically normalized, we can use the freedom to shift the energy levels by some
offset $s\in\R$, i.e., $E'_k:=E_k+s$, $E':=E+s$. The resulting distribution $\lambda'_i:=\frac{E'}{|A|}\cdot
\frac 1 {E'_i}$ is normalized, i.e., $\sum_i \lambda'_i=1$, if and only if
\[
   E'=\left(\frac 1 {|A|} \sum_{k=1}^{|A|} \frac 1 {E'_k}\right)^{-1}=:E'_H,
\]
that is, if the offset is shifted such that the new energy value $E'$ equals the harmonic mean energy $E'_H$.
We have thus reproduced the canonical density matrix $\rho_c$ of Theorem~\ref{MainTheorem3}. Moreover,
the calculation above shows that the occurrence of the
harmonic mean energy is very natural and not a technical artifact of our proof. (In Section~\ref{SecGaussian},
we give a method for numerical sampling of the energy manifold, and there, the harmonic mean energy
will appear in a natural way as well.)

\section{Proof of main theorem}
\label{SecConcentration}
Before proving the main theorems, we fix some notation that will be useful for the proof.
Following the lines of Gromov~\cite{Gromov}, we define a \emph{metric measure space} $X$ to be a
separable complete metric space with a finite Borel measure $\mu$, i.e., $X=(X,{\rm dist},\mu)$.
(In fact, one could more generally consider \emph{Polish spaces} with a $\sigma$--finite Borel measure, as described by Gromov.)
In this paper, we will only consider the two cases that $X$ is the energy manifold $M_E$, or a full
ellipsoid $N$, both equipped with the obvious geometric measure and the metric ${\rm dist}$ that
is induced by the surrounding Euclidean space $\R^{2n}\simeq\C^n$ (for more details, see Section~\ref{SecConcentration}).

We denote the $(n-1)$-sphere by $S^{n-1}$, and the $n$-ball is denoted $B^n$, i.e.,
\begin{eqnarray*}
   S^{n-1}&:=&\left\{x\in\R^n \,\,|\,\, \|x\|=1\right\},\\
   B^n&:=&\left\{x\in\R^n \,\,|\,\, \|x\|\leq 1\right\}.
\end{eqnarray*}
The symbol $\partial$ is used for the topological boundary, for example, $\partial B^n = S^{n-1}$.

We denote the $k$-dimensional (Hausdorff) volume measure on $k$-dimensional submanifolds $M\subset \R^n$ by $\mu_k$.
In case that $\mu_k(M)<\infty$, we write $\mu_M$ for the normalized measure on $M$, i.e.,
\begin{equation*}
	\mu_M(X):=\frac{\mu_k(X)}{\mu_k(M)} 
\end{equation*}	
for Borel subsets $X\subset M$. Sometimes we consider
subsets that are not actually submanifolds, but are turned into submanifolds in an obvious way.
For example, the Ball $B^n$ is itself a metric measure space, but not a submanifold, since it is
not open. However, its interior is a submanifold of $\R^n$, and
\begin{equation*}
	\mu_n(B^n)=\frac{\pi^{\frac n 2}}
{\Gamma\left(1+\frac n 2\right)}, 
\end{equation*}
while $\mu_{B^n}(B^n)=1$. Expectation values with respect to $\mu_M$
are denoted $\mathbb{E}_M$.

Given some Hamiltonian $H=H^\dagger$ on $\C^m$, we would like to prove concentration of measure
for the ``mean energy ensemble''
\[
   M_E:=\left\{\psi\in\C^m \,\,|\,\, \langle \psi|H|\psi\rangle=E,\quad \|\psi\|=1\right\}.
\]
To relate our discussion to real-valued geometry, we work instead in $\R^n$ with $n=2m$.
That is, doubling all energy eigenvalues of $H$ to get the energy levels $E_1,\ldots,E_n$,
and slightly abusing notation, we can write
\[
   M_E=\left\{x\in\R^n\,\,\left|\,\, \sum_{k=1}^n x_k^2 E_k = E, \quad \sum_{k=1}^n x_k^2=1\right.\right\}.
\]
Geometrically, $M_E$ is the intersection of the unit sphere with an ellipsoid (given by the energy condition).
The action of shifting all energies by some offset (as postulated in Theorem~\ref{MainTheorem1}) alters the
ellipsoid, but not the ellipsoid's intersection with the sphere; it leaves $M_E$ invariant.
It is interesting to note that the corresponding full ellipsoid's volume turns out to be minimal
exactly if the energy shift is tuned such that $E'=E_H'$, i.e., for the harmonic mean energy shift
close to the one which is postulated in Theorem~\ref{MainTheorem1}.

We would like to prove measure concentration for the algebraic variety $M_E$ in ordinary Euclidean space.
Introducing a function $f:\R^n\to\R^2$ via
\[
   f(x):=\left( \sum_{k=1}^n x_k^2 \quad,\quad \sum_{k=1}^n x_k^2 E_k \right)^T,
\]
we can write 
\begin{equation*}
	M_E=f^{-1}\left(\begin{array}{c} 1 \\ E \end{array}\right). 
\end{equation*}
If $E$ is any energy value
with $\min_k E_k < E < \max_k E_k$ and $E\neq E_k$ for all $k$, then the differential $df$ has full rank on
all of $M_E$, such that $M_E$ is a proper submanifold of $\R^n$ of codimension $2$.

If $E=E_k$ for some $k$ (but $E\neq \min_k E_k$ and $E\neq \max_k E_k$),
the eigenvectors corresponding to this energy value are singular points of $f$.
Still, removing those eigenvectors from $M_E$, we get a valid submanifold $\tilde M_E$ of $\R^n$ of codimension $2$.
Since $\tilde M_E$ and $M_E$ agree up to a set of measure zero, we will drop the tilde in the following,
and simply write $M_E$ for the manifold with eigenvectors removed. If we treat $M_E$ as a metric measure space,
we include the eigenvectors in its definition to have a complete metric space.

As a submanifold of $\R^n$, the sets $M_E$ carry a natural geometric volume measure. Since every $M_E$ is
a compact submanifold, it makes sense to talk about the normalized measure $\mu_{M_E}$ on $M_E$,
and to ask whether this measure exhibits a concentration of measure phenomenon. Our main proof strategy
to answer this question in the positive is due to Gromov~\cite{Gromov}.
To explain this strategy, we introduce the notion of a ``typical
submanifold''. Suppose that $N$ is a metric measure space, and $M\subset N$ is a subset (say, a submanifold) which
is itself a metric measure space. Then we say that ``$M$ is typical in $N$'' if $M$ has small codimension, and if a small neighborhood
of $M$ covers almost all of $N$. That is, $\mu_N(U_\varepsilon(M))\approx 1$ already for small $\varepsilon$.

For example, given an $n$-dimensional sphere $N=S^n$, any equator $M$ (which is itself an $(n-1)$-dimensional sphere) is typical
in $N$; this is just L\'evy's Lemma. On the other hand, a polar cap with angle $\theta$ is only typical in $S^n$ if
$\theta\approx\frac\pi 2$.

Gromov's idea can now be explained as follows:
\begin{itemize}
\item[] If $N$ is a metric measure space that shows concentration of measure, and if $M$ is a
typical submanifold of $N$, then $M$ shows measure concentration as well: it ``inherits'' concentration
of measure from $N$.
\end{itemize}
The intuitive reason why this idea works is as follows. Consider the behavior of Lipschitz-continuous functions on $M\subset N$.
By continuity, those functions do not change much if we turn to an $\varepsilon$-neighborhood of $M$; but then,
since $M$ is assumed to be typical in $N$, we already obtain almost all of $N$, and so the behavior of the functions
on $M$ will be similar to that on $N$ -- in particular, expectation values will be similar, and the concentration of
measure phenomenon will occur in $M$ if it occurs in $N$. However, Gromov seems to explore this idea in his book only for the case
that $N$ is a sphere.

In our case, we have to find a submanifold $N\subset\R^n$ which is itself subject to the concentration
of measure phenomenon, such that $M_E\subset N$ holds
and such that $M_E$ is typical in $N$. A first obvious guess is to use the sphere itself; clearly, $M_E$ is a subset
of the sphere $S^{n-1}\subset\R^n$, and we have concentration of measure on the sphere by L\'evy's Lemma.
However, $M_E$ can only be typical in $S^{n-1}$ if the energy $E$ is close to the ``typical'' value $\mathbb{E}_{S^{n-1}} \sum_{k=1}^{n}x_k^2 E_k$,
which turns out to be the mean value $\frac 1 n(E_1+E_2+\ldots+E_n)$. Since we definitely want to consider
different energies far away form the mean energy, $N=S^{n-1}$ is not a useful choice.
Instead, it will turn out that we can choose $N$ to be a \emph{full ellipsoid} with appropriate equatorial
radii (in fact, $N$ will be the slightly enlarged full energy ellipsoid for an appropriate energy shift).

Our main tool from integral geometry is the Crofton formula~\cite{Santalo,Tasaki}. It expresses the volume of
a submanifold of $\R^n$ in terms of the average volume of intersection of that manifold with random hyperplanes.

\begin{lemma}[The Crofton formula~\cite{Santalo}]
\label{LemCrofton}
Let $M$ be a $q$-dimensional submanifold of $\R^n$. Consider the invariant
measure $dL_r$ on the planes of dimension $r$ in $\R^n$. If $r+q\geq n$,
\[
   \int_{L_r} \mu_{r+q-n}(M\cap L_r)\, dL_r = \sigma(q,r,n) \mu_q(M),
\]
where $\displaystyle\sigma(q,r,n)=\frac{O_{r+q-n}\cdot\prod_{i=n-r}^n O_i}{O_q\cdot\prod_{i=0}^r O_i}$, and
$O_n:=\frac{2 \pi^{\frac{n+1}2}}{\Gamma\left( \frac{n+1}2\right)}$ denotes the surface area of the $n$-sphere.
\end{lemma}
Note that the Crofton formula is formulated in Ref.\ \cite[(14.69)]{Santalo} only for the case that $M$ is a
\emph{compact} submanifold, but the proof remains valid also in the case that $M$ is not compact.
This observation is also expressed in Ref.\ \cite{Tasaki}.

For the details of the definition of the invariant measure $dL_R$, see Ref.\ \cite{Santalo}. In short, the Lie group $\mathfrak{M}$
of all motions (translations and rotations) in $\R^n$ possesses the closed subgroup $\mathfrak{H}_r$ of motions leaving
a fixed $r$-dimensional plane $L_r^0$ invariant. Then, there is a one-to-one correspondence between the set of $r$-planes
in $\R^n$ and the homogeneous space $\mathfrak{M}/{\mathfrak{H}_r}$. Since both $\mathfrak{M}$ and $\mathfrak{H}_r$ are
unimodular, $\mathfrak{M}/{\mathfrak{H}_r}$ possesses an invariant density $dL_r$ which can then be interpreted as a density
on the $r$-planes in $\R^n$.

Note that this measure is defined only up to some multiplicative constant -- different authors use different
normalizations (cf. Ref.\ \cite{SchneiderWeil}), which gives different constants $\tilde\sigma(q,r,n)$ instead of $\sigma(q,r,n)$
in Lemma~\ref{LemCrofton}. However, for fixed $l$ and $n$, we always have
\begin{equation*}
	\displaystyle \frac{\sigma(k,l,n)}{\sigma(k',l,n)}=\frac{\tilde\sigma(k,l,n)}{\tilde\sigma(k',l,n)}, 
\end{equation*}	
	and it is only those ratios that
are relevant for the calculations. If $r+q=n$, then $\mu_{r+q-n}(X)=\mu_0(X)$ equals the number of points in the set $X$.
A useful possible expression for the constants is~\cite{SchneiderWeil}
\[
   \tilde \sigma(q,r,n)=\frac{\Gamma\left(\frac{q+1} 2 \right)
   \Gamma\left(\frac{r+1} 2 \right)}{\Gamma\left(\frac{n+1} 2 \right) \Gamma\left(\frac{q+r-n+1}2\right)}.
\]

The Crofton formula will be useful in the following lemma, which
gives a lower bound on the measure of $\varepsilon$-neighborhoods of subsets of $M_E$. Given any
subset $X\subset\R^n$ (say, any curve), directly estimating $\mu_n(U_\varepsilon(X))$ seems difficult -- if
the curve intersects itself many times, the neighborhood can be almost arbitrarily small. On the other
hand, the Crofton formula takes into account how ``meandering'' subsets $X$ are, by counting the
number of intersections with hyperplanes.
\begin{lemma}[Measure of neighborhood]
\label{LemMeasureNeighborhood}
For every open subset $X\subset M_E\subset\R^n$ with $n\geq 3$ and $\varepsilon>0$, we have
\[
   \mu_n(U_\varepsilon(X))\geq \frac{\pi\varepsilon^2}{4(n-1)} \mu_{n-2}(X),
\]
where $U_\varepsilon(X)$ denotes the open $\varepsilon$-neighborhood in $\R^n$ of $X$.
\end{lemma}
\proof
If $L_2$ is any $2$-dimensional plane in $\R^n$, then $M_E\cap L_2=(\partial N\cap L_2)\cap (S^n\cap L_2)$, where $\partial N$
is the surface of an ellipsoid. If the intersection is not empty, then $S^n\cap L_2$ is a circle,
and $\partial N\cap L_2$ is a compact quadratic curve, hence an ellipse. By the Crofton formula,
\[
   \int_{L_2} \mu_0(M_E\cap L_2)\, dL_2 = \sigma(n-2,2,n)\mu_{n-2}(M_E)<\infty,
\]
so the set of planes $L_2$ with $\mu_0(M_E\cap L_2)=\#(M_E\cap L_2)=\infty$ has measure zero, and we can
ignore them. Let now $L_X$ be the set of all planes $L_2$ such that $X\cap L_2$ is a finite non-empty set.
Since a circle and an ellipse in the plane intersect in at most four points if they are not equal,
we have $\#(X\cap L_2)\leq 4$ for all $L_2\in L_X$. Hence, the Crofton formula yields
\begin{eqnarray*}
\sigma(n-2,2,n)\mu_{n-2}(X)&=&\int_{L_2} \mu_0(X\cap L_2)\,dL_2 \\
&=& \int_{L_X} \#(X\cap L_2)\, dL_2\\
&\leq& 4 \int_{L_X}\,dL_2.
\end{eqnarray*}
Using Crofton's formula for the $n$-dimensional submanifold $U_\varepsilon(X)$, we get on the other hand
\begin{eqnarray*}
   \sigma(n,2,n)\mu_n(U_\varepsilon(X))&=&\int_{L_2} \mu_2(U_\varepsilon(X)\cap L_2)\, dL_2 \\
   &\geq &\int_{L_X} \mu_2(U_\varepsilon(X)\cap L_2)\, dL_2\\
   &\geq& \pi\varepsilon^2 \int_{L_X}\, dL_2,
\end{eqnarray*}
since every plane that intersects $X$ intersects $U_\varepsilon(X)$ in at least a disc of radius $\varepsilon$.
Combining both inequalities, the claim follows.
\qed

To deal with concentration of measure on submanifolds, we need to introduce some additional
notions of the theory of measure concentration; they all can be found in the book by Gromov~\cite{Gromov},
and also, e.g., in Ref.~\cite{Funano} with a few errors corrected.

Let $(Y,\nu)$ be a metric measure space with $0<m:=\nu(Y)<\infty$, then the ``partial diameter'' ${\rm diam}$ is defined by
\[
   {\rm diam}(\nu,m-\kappa):=\inf\{{\rm diam}(Y_0)\,\,|\,\, Y_0\subset Y,\nu(Y_0)\geq m-\kappa\},
\]
that is, the smallest diameter of any Borel subset with measure larger than $m-\kappa$.
If $(X,\mu)$ is a metric measure space with $m:=\mu(X)<\infty$,
the {\em observable diameter} $\ObsDiam$ is defined as
\begin{eqnarray*}
   \ObsDiam(X,\kappa)&:=& \sup\left\{ {\rm diam}(\mu\circ f^{-1},m-\kappa)\right.\,\,|\\
    &&\left. f:X\to\R \mbox{ is 1-Lipschitz}\right\},
\end{eqnarray*}
where $\mu\circ f^{-1}$ is the push-forward measure on $f(X)\subset \R$.
In the special case that $\mu=\mu_X$, i.e., if $\mu$ is normalized such that $m=1$, this definition implies
\begin{eqnarray}
\ObsDiam(X,\kappa)\leq D \quad\Leftrightarrow\quad \forall f:X\to\R\mbox{ 1-Lipschitz }\nonumber\\
\forall \varepsilon>0 \exists Y_0\subset f(X)\subset \R
 \mbox{ with }\mu_X\left(f^{-1}(Y_0)\right)\geq 1-\kappa\nonumber\\
 \mbox{ and }{\rm diam}(Y_0)\leq D+\varepsilon.\qquad\strut\label{EqCharacterizationObsDiam}
\end{eqnarray}
This shows that small $\ObsDiam$ amounts to a large amount of measure concentration -- in fact,
one can conversely infer that $\ObsDiam(X,\kappa)\leq D$ for $D>0,0<\kappa<\frac 1 2$ implies
that all $\lambda$-Lipschitz maps $f:X\to\R$ satisfy
\begin{equation}
   \mu_X\left\{x\in X\,\,:\,\, |f(x)-m_X f |\geq \lambda D\right\}\leq \kappa,
   \label{EqObsDiamConsequence}
\end{equation}
where $m_X f$ is the median of $f$ on $X$, see Ref.\ \cite{Gromov} and, for a proof, Ref.\ \cite[Lemma 2.3]{Funano2}.
(Replacing $\kappa$ by $2\kappa$ on the right-hand side removes the restriction $\kappa<\frac 1 2$.)
Another useful notion is the \emph{separation distance}
\begin{eqnarray*}
   \Sep(X;\kappa_0,\ldots,\kappa_N)&:=&\sup\{ \delta\,\,|\,\, \exists X_i\subset X:\, \mu(X_i)\geq\kappa_i,\\
   &&{\rm dist}(X_i,X_j)\geq \delta,\enspace X_i \mbox{ open}\}.
\end{eqnarray*}
In Ref.\ \cite{Gromov}, arbitrary Borel sets are allowed in the definition of the separation distance; here,
we only use open sets, since they are submanifolds and hence subject to the Crofton formula.
In the most important case of two parameters, the equation $\Sep(X,\kappa,\kappa)=D$
implies that for every $\varepsilon>0$ there are open subsets $X_1,X_2\subset X$ with $\mu(X_1)\geq\kappa$
and $\mu(X_2)\geq \kappa$ such that ${\rm dist}(X_1,X_2)\geq D-\varepsilon$.

In the following, we need an inequality relating separation distance and observable diameter.
It has first been stated in Ref.~\cite{Gromov}, and a small lapse has been corrected in Ref.~\cite{Funano}.
Since the notation of Ref.~\cite{Funano} differs significantly from the notation used here,
we give a proof in order to keep the presentation self-contained.

\begin{lemma}[Observable diameter]\label{LemObsDiamSep}
For every metric measure space $X$, and
for every $\kappa>\kappa'>0$, it holds
\[
   \ObsDiam(X,2\kappa)\leq\Sep(X;\kappa,\kappa)\leq \ObsDiam(X,\kappa').
\]
\end{lemma}
\proof
Suppose that $\ObsDiam(X;\kappa')\leq D$. Let $X_1\subset X$ be an open set with
$\mu_X(X_1)\geq \kappa$, and define a function $f:X\to\R$ via
\[
   f(x):={\rm dist}(x,X_1)=\inf_{y\in X_1} {\rm dist}(x,y),
\]
where $d$ denotes the metric on $X$. It follows from the triangle inequality that $f$
is continuous with Lipschitz constant $1$. Let $\varepsilon>0$.
According to~(\ref{EqCharacterizationObsDiam}), there is a subset $Y_0\subset\R$ with
${\rm diam}(Y_0)\leq D+\varepsilon$ such that $\mu_X\{x\in X\,\,|\,\, f(x)\in Y_0\}\geq 1-\kappa'$.
Since $1-\kappa'+\kappa>1$, it follows that $X_1\cap f^{-1}(Y_0)\neq \emptyset$,
hence we have $0\in Y_0$.
Similarly, if $X_2\subset X$ is another open set with $\mu_X(X_2)\geq\kappa$, it
follows that $X_2\cap f^{-1}(Y_0)\neq \emptyset$, so there is some $x\in X_2$ with
$f(x)\in Y_0$. Thus,
\[
   {\rm dist}(X_1,X_2)=\inf_{x\in X_2} f(x)\leq \sup_{y\in Y_0} y ={\rm diam}(Y_0)\leq D+\varepsilon.
\]
Since $\varepsilon>0$ was arbitrary, it follows that ${\rm dist}(X_1,X_2)\leq D$, and thus $\Sep(X;\kappa,\kappa)\leq D$.
This proves the second inequality.

To prove the first inequality, suppose that $\ObsDiam(X,2\kappa)\geq D$. This means that there
exists a $1$-Lipschitz function $f:X\to\R$ such that for all $Y_0\subset\R$ with
\begin{equation*}
	\mu_X\{x\in X\,\,|\,\, f(x)\in Y_0\}\geq 1-2\kappa 
\end{equation*}
it holds ${\rm diam}(Y_0)\geq D$.
Clearly, the function $M(a):=\mu_X\left(f^{-1}\left((-\infty, a]\right)\right)$ is
increasing in $a\in\R$. Let $a_0:=\inf\{a\in\R\,\,|\,\, M(a)\geq\kappa\}$, then
\begin{equation*}
	\mu_X\left( f^{-1}\left((-\infty, a_0]\right)\cup f^{-1}\left((a_0,a]\right)\right)\geq\kappa
\end{equation*}	
for all $a>a_0$. Since every finite Borel measure on a Polish space is regular~\cite[Ulam Theorem]{Elstrodt},
it follows that $\mu_X\left(f^{-1}\left((-\infty,a_0]\right)\right)\geq\kappa$.
Similarly, define $N(b):=\mu_X\left(f^{-1}\left([b,\infty)\right)\right)$, and let 
\begin{equation*}
b_0:=\sup\{b\in\R\,\,|\,\, N(b)\geq\kappa\}.
\end{equation*}	
An analogous argument shows that $\mu_X\left(f^{-1}\left([b_0,\infty)\right)\right)\geq\kappa$. Moreover,
$\mu_X\left(f^{-1}\left((a,b)\right)\right)\geq 1-2\kappa$ if $a<a_0$ and $b>b_0$. Due to the regularity of $\mu_X$,
we conclude that $\mu_X\left(f^{-1}\left(Y_0\right)\right)\geq 1-2\kappa$ for $Y_0:=[a_0,b_0]$.
Setting $Y_1:=(-\infty,a_0]$ and $Y_2:=[b_0,\infty)$, we get ${\rm dist}(Y_1,Y_2)\geq {\rm diam}(Y_0)\geq D$.
Letting $X_i:=f^{-1}(Y_i)$ for $i=1,2$, it follows from the Lipschitz continuity of $f$ that
${\rm dist}(X_1,X_2)\geq D$, and $\mu_X(X_i)\geq \kappa$. Now if $\varepsilon>0$ is arbitrary,
the open sets $\tilde X_i:=U_\varepsilon(X_i)$ have measure $\mu_X(\tilde X_i)\geq\kappa$,
and ${\rm dist}(\tilde X_1,\tilde X_2)\geq D-2\varepsilon$. This proves that $\Sep(X;\kappa,\kappa)\geq D$.
\qed

Now we can formulate a lemma on how $M_E$ inherits measure concentration from surrounding bodies:
\begin{lemma}[Measure concentration on $M_E$ from that of surrounding body]
\label{LemConSurroundingBody}
Let $\varepsilon>0$, and let $N\subset\R^n$ be a metric measure space
such that $U_\varepsilon(M_E)\subset N$. Then, for all $\kappa>0$
\[
   \ObsDiam(M_E,2\kappa)\leq 2\varepsilon+
   \ObsDiam\left(N,\frac{\mu_{n-2}(M_E)\pi\varepsilon^2\kappa}{\mu_n(N) 4 (n-1)}\right).
\]
\end{lemma}
\proof
Abbreviate $X:=M_E$.
Suppose that $\Sep(X;\kappa,\kappa)=D$, that is, for every $\delta>0$, there are open subsets $X_1,X_2\subset X$
such that $\mu_X(X_1)\geq\kappa$ and
$\mu_X(X_2)\geq \kappa$ and ${\rm dist}(X_1,X_2)\geq D-\delta$. Let $\tilde X_i:=U_\varepsilon(X_i)\subset N$.
Using Lemma~\ref{LemMeasureNeighborhood}, we get
\begin{eqnarray*}
   \mu_N(\tilde X_i)&=&\frac{\mu_n(\tilde X_i)}{\mu_n(N)} \geq \frac 1 {\mu_n(N)}\frac{\pi\varepsilon^2}{4(n-1)}\mu_{n-2}(X_i)\\
   &=&\frac 1 {\mu_n(N)}\frac{\pi \varepsilon^2}{4(n-1)} \mu_X(X_i)\mu_{n-2}(X)\\
   &\geq& \frac {\mu_{n-2}(X)} {\mu_n(N)}
   \frac{\pi\varepsilon^2}{4(n-1)} \kappa=:\kappa_\varepsilon.
\end{eqnarray*}
Since also ${\rm dist}(\tilde X_1,\tilde X_2)\geq D-\delta-2\varepsilon$, we get $\Sep(N;\kappa_\varepsilon,\kappa_\varepsilon)\geq D-2\varepsilon$.
All in all, we have shown that
\[
   \Sep(X;\kappa,\kappa)\leq 2\varepsilon+\Sep(N;\kappa_\varepsilon,\kappa_\varepsilon).
\]
Lemma~\ref{LemObsDiamSep} yields the chain of inequalities
\begin{eqnarray*}
\ObsDiam(X,2\kappa)&\leq& \Sep(X;\kappa,\kappa)\\
&\leq& \Sep(N;\kappa_\varepsilon,\kappa_\varepsilon)+2\varepsilon \\
&\leq& \ObsDiam(N,\kappa'_\varepsilon)+2\varepsilon
\end{eqnarray*}
for every $\kappa'_\varepsilon<\kappa_\varepsilon$.\qed

Our goal in the following will thus be to find a good $n$-dimensional body $N$ with small observable
diameter and $U_\varepsilon(M_E)\subset N$ such that the ratio $\frac{\mu_{n-2}(M_E)}{\mu_n(N)}$ appearing
in the previous lemma is not too small. To this end, note that
\[
   \mu_{n-2}(M_E)=\mu_{n-2}(\partial(S^{n-1}\cap N_E)),
\]
where $N_E=\left\{x\in\R^n\,\,|\,\, \sum_{k=1}^n E_k x_k^2\leq E\right\}$ is the full energy ellipsoid.
We would like to relate $\mu_{n-2}(\partial(S^{n-1}\cap N_E))$ to $\mu_{n-1}(S^{n-1}\cap N_E)$. For this
purpose, the following isoperimetric inequality will be useful.

\begin{lemma}[An isoperimetric inequality]
\label{LemIsoperimetry2}
Let $n\geq 3$, and let $B\subset S^{n-1}\subset \R^n$ be a Borel subset which covers at most half of the sphere,
i.e., $\mu_{S^{n-1}}(B)\leq\frac 1 2$. Then,
\[
   \frac{\mu_{n-2}(\partial B)}{\mu_{n-1}(B)}> \frac 1 2 \sqrt{n}.
\]
\end{lemma}
\proof We use the isoperimetric inequality on the sphere~\cite[Appendix I]{Ledoux,MilmanSchechtman}: 
\emph{Among all Borel sets in $S^{n-1}$ with fixed volume, the minimal volume of the boundary is
assumed by a round ball.} Thus, let $C_t^{n-1}\subset S^{n-1}$ be a polar cap (i.e., round ball)
with corresponding angle $0< t\leq\frac\pi 2$ such that $\mu_{n-1}(C_t^{n-1})=\mu_{n-1}(B)$. By the isoperimetric
inequality,
\begin{eqnarray*}
   \frac{\mu_{n-2}(\partial B)}{\mu_{n-1}(B)}&\geq& \frac{\mu_{n-2}(\partial C_t^{n-1})}{\mu_{n-1}(C_t^{n-1})}\\
   &=&\frac{\mu_{n-2}(S^{n-2})\cdot \sin^{n-2} t}{\mu_{n-2}(S^{n-2})\cdot \int_0^t \sin^{n-2}\theta\, d\theta}\\
   &=&\frac{\sin^{n-2} t}{\int_0^t \sin^{n-2}\theta\,d\theta}=:f_n(t).
\end{eqnarray*}
For $t\in[0,\pi/2)$, let $h_n(t):=(n-2)\int_0^t \sin^{n-2}\theta\, d\theta - \frac{\sin^{n-1} t}{\cos t}$,
then $h'_n(t)=-\sin^{n-2}t(1+\tan^2 t)\leq 0$, so $h_n$ is decreasing. Since $h_n(0)=0$, it follows that
$h_n(t)\leq 0$ for all $t\in[0,\pi/2)$. Multiplying the corresponding equation with the non-negative expression
$\frac{\cos t}{\sin t\,\int_0^t\ldots}$ gives for $0<t<\frac \pi 2$
\[
   (n-2)\frac{\cos t}{\sin t}-\frac{\sin^{n-2} t}{\int_0^t \sin^{n-2}\theta\,d\theta}\leq 0.
\]
But the left-hand side is exactly $(\ln f_n(t))'$, such that $\ln f_n(t)$ is decreasing, and hence
$f_n(t)$ is decreasing, so
\[
   f_n(t)\geq f_n\left(\frac\pi 2\right)=\frac{2\Gamma\left(\frac n 2\right)}{\sqrt{\pi}\Gamma\left(\frac{n-1}2\right)},
\]
and this expression is larger than $\frac 1 2 \sqrt{n}$ if $n\geq 3$ (it grows asymptotically like $\sqrt{\frac 2 \pi}\sqrt{n}$).
\qed

To deal with ellipsoids, we need some results on expectation values of certain functions.
\begin{lemma}[Ellipsoidal expectation values]
\label{LemEllipsoidExp}
Let $N\subset\R^n$ be the full ellipsoid with equatorial radii $\{a_k\}_{k=1}^n$, $a_k>0$. Then we have
the following expectation values with respect to the geometric measure in $N$:
\begin{eqnarray*}
   \mathbb{E}_N \|x\|^2 &=&\frac{a_1^2+a_2^2+\ldots + a_n^2}{n+2},\\
   \mathbb{E}_N \|x\|^4 &=&\frac{2\sum_{k=1}^n a_k^4 + \left(\sum_{k=1}^n a_k^2\right)^2}{(n+2)(n+4)}.
\end{eqnarray*}
\end{lemma}
\proof
First, we use a linear transformation to reduce the expectation value calculations to integrals
on the ball $B^n=\{x\in\R^n:\|x\|\leq 1\}$. That is, let $\Phi(x):={\rm diag}(a_1,\ldots,a_n)x$ for $x\in\R^n$,
then by the transformation formula for integrals, we have
\[
   \int_N f(x)dx=\int_{B^n}f(\Phi(x))|\det D\Phi(x)|dx,
\]
where $\det D\Phi(x)=a_1 a_2\ldots a_n$. The only other non-trivial ingredient are the spherical integrals
\begin{eqnarray}
   \int_{B^n} x_1^2 dx&=& \frac{\mu_n(B^n)}{n+2},\label{EqIntx2bn}\\
   \int_{B^n} x_1^4 dx &=& \frac{3 \mu_n(B^n)}{(n+2)(n+4)},\nonumber\\
   \int_{B^n} x_1^2 x_2^2 dx &=& \frac{\mu_n(B^n)}{(n+2)(n+4)}.\nonumber
\end{eqnarray}
These formulas can be proved directly by applying hyperspherical coordinates, cf. Ref.\ \cite{hyperspherical}.
\qed

Now we are ready to estimate the crucial expression $\frac{\mu_{n-2}(M_E)}{\mu_n(N)}$ for surrounding ellipsoids $N$.

\begin{lemma}[Ratio of $M_E$ and the surrounding ellipsoid]
\label{LemCrucialEstimate}
$\strut$\linebreak
Let $n\geq 3$, and 
\[
	E_H:=\left(\frac 1 n \sum_{k=1}^n \frac 1 {E_k}\right)^{-1}
\]
be the
harmonic mean energy corresponding to the energy levels $E_k>0$, $k=1,\ldots,n$. Suppose that
\[
   E=\frac{n+2} n (1+\delta)E_H
\]
with some $\delta>0$ such that the energy manifold $M_E$ is not empty. Moreover, suppose that $E$ is
less than the median of $\sum_k E_k x_k^2$ on the sphere $S^{n-1}$.
Let $N$ be a full ellipsoid with equatorial radii $\{a_k\}_{k=1}^n$, $a_k>0$, i.e.,
\[
    N:=\left\{x\in\R^n\,\,\left|\,\, \sum_{k=1}^n \frac{x_k^2}{a_k^2} \leq 1\right.\right\}.
\]
Then, we have
\begin{eqnarray*}
   \frac{\mu_{n-2}(M_E)}{\mu_n(N)}&>& \frac 1 2 \cdot \frac{n^{\frac 3 2}}{(1+2\delta)^{\frac n 2}-1}
   \prod_{k=1}^n \left({\frac{E}{a_k^2 E_k}}\right)^\frac12 \times \\
   &&\times \left(1-\frac{2 E^2}{\delta^2(n+2)(n+4)} \sum_{k=1}^n \frac 1 {E_k^2}\right).
\end{eqnarray*}
\end{lemma}
\proof
According to the isoperimetric inequality from Lemma~\ref{LemIsoperimetry2}, we have (with $N_E$
the energy ellipsoid as defined directly above that lemma),
\[
   \mu_{n-2}(M_E)=\mu_{n-2}(\partial(S^{n-1}\cap N_E))>\frac {\sqrt{n}} 2\mu_{n-1}(S^{n-1}\cap N_E).
\]
Hence it remains to lower-bound $\mu_{n-1}(S^{n-1}\cap N_E)$; this is exactly the Haar measure probability
${\rm Prob}\left\{ \langle\psi|H|\psi\rangle\leq E\right\}$. In principle, this probability can be
computed exactly: using the volume-preserving map from the unit sphere to the probability simplex~\cite{BengtssonZyk},
this probability equals the ratio of the volumes of the two bodies that originate from intersecting the
simplex with a hyperplane. This ratio has been computed in Ref.\ \cite{Dempster}, and the result is
$\sum_{k:E_k\leq E} \prod_{i\neq k} \frac{E-E_k}{E_i-E_k}$ for $E\leq E_{\text{max}}$. Unfortunately, the
result is only valid in the non-degenerate case; moreover, despite its simple form, it is hard to
estimate that value in a way which is useful in the current calculation. Thus, we instead use
a different approach which is based on geometry of the ellipsoid.
Let $E(x):=\sum_{k=1}^n E_k x_k^2$, and
let $S_r^{n-1}$ be the sphere of radius $r$ in $\R^n$. First we show the inequality
\begin{equation}
   \mu_{n-1}(S_{\lambda r}^{n-1}\cap N_E)\geq \lambda^{n-1}\mu_{n-1}(S_r^{n-1}\cap N_E)
   \label{EqIneq}
\end{equation}
for any $\lambda\in(0,1)$. This can be seen as follows: let $x\in S_r^{n-1}\cap N_E$ such that $\|x\|=r$
and $E(x)\leq E$. Then, $\|\lambda x\|=\lambda\|x\|$ and $E(\lambda x)=\lambda^2 E(x)<E$, such
that $x\in S_{\lambda r}^{n-1}\cap N_E$. Hence $\lambda(S_r^{n-1}\cap N_E)\subset S_{\lambda r}^{n-1}\cap N_E$, and so
$\mu_{n-1}(S_{\lambda r}^{n-1}\cap N_E)$ is lower-bounded by $\mu_{n-1}(\lambda(S_r^{n-1}\cap N_E))$, which
equals $\lambda^{n-1}\mu_{n-1}(S_r^{n-1}\cap N_E)$.

Let $B_{a,b}^n$ be the set of vectors in $\R^n$ with norm between $a$ and $b$.
With the help of Inequality~(\ref{EqIneq}), we get
\begin{eqnarray*}
   \mu_n(N_E\cap B_{1,\sqrt{1+2\delta}}^n)&=& \int_1^{\sqrt{1+2\delta}} \mu_{n-1}(S_r^{n-1}\cap N_E)\, dr\\
   &\leq&  \int_1^{\sqrt{1+2\delta}} r^{n-1} \mu_{n-1}(S^{n-1}\cap N_E)\, dr\\
   &=& \mu_{n-1}(S^{n-1}\cap N_E)\cdot \frac{(1+2\delta)^{\frac n 2}-1}{n}.
\end{eqnarray*}
Thus, we have reduced the problem to finding a lower bound on $\mu_n(N_E\cap B_{1,\sqrt{1+2\delta}})$.
Indeed, applying Lemma~\ref{LemEllipsoidExp} to the assumptions of this lemma, we see that
the expectation value of $\|x\|^2$ with respect to the geometric measure in $N_E$ is exactly
\[
   \mathbb{E}_{N_E} \|x\|^2 = 1+\delta,
\]
so we can indeed expect that much of the weight of $N_E$ is contained in $B^n_{1,\sqrt{1+2\delta}}$.
To prove this, we use the Chebyshev inequality. Let $\sigma^2$ be the variance of $\|x\|^2$ with
respect to the geometric measure on $N_E$, then it is easy to see that
\[
   \sigma^2=\mathbb{E}_{N_E}\|x\|^4 - \left(\mathbb{E}_{N_E}\|x\|^2\right)^2
   \leq \frac{2 E^2}{(n+2)(n+4)} \sum_{k=1}^n \frac 1 {E_k^2},
\]
and the probability that a point in $N_E$ is not contained in $B^n_{1,\sqrt{1+2\delta}}$ is upper-bounded
by $\frac {\sigma^2}{\delta^2}$. Hence
\begin{eqnarray*}
   \mu_n(N_E\cap B^n_{1,\sqrt{1+2\delta}})&\geq& \left(1-\frac{2 E^2}{\delta^2(n+2)(n+4)} \sum_{k=1}^n \frac 1 {E_k^2}\right)\times\\
   &&\times \mu_n(N_E).
\end{eqnarray*}
The claim follows from substituting explicit expressions for $\mu_n(N_E)$ and $\mu_n(N)$.
\qed

Since we want to show that our energy submanifold $M_E$ inherits measure concentration from an ellipsoid,
we first have to prove concentration of measure for ellipsoids:
\begin{lemma}[Measure concentration for ellipsoids]
\label{LemConEll}
Let $N\subset\R^n$ be the full ellipsoid with equatorial radii $\{a_k\}_{k=1}^n$, $a_k>0$. Then, for
every $\kappa>0$, we have
\[
   \ObsDiam(N,\kappa)\leq \frac{4a}{\sqrt{n}}\sqrt{\ln\frac 4 \kappa},
\]
where $a:=\max_k a_k$.
\end{lemma}
\proof
In accordance with Refs.~\cite{Barvinok,Ledoux},
for a (convex) body $K\subset\R^n$ with surface $S:=\partial K$, we say that \emph{$K$ is strictly convex} if for
every $\varepsilon>0$ there exists some $\delta=\delta(\varepsilon)>0$ such that $x,y\in S$ and $\|x-y\|\geq \varepsilon$
implies $(x+y)/2\in (1-\delta)K$. The unit ball $B^n$ is strictly convex, and one can choose
\[
   \delta(\varepsilon)=1-\left({1-\frac{\varepsilon^2} 4}\right)^\frac12 \geq\frac{\varepsilon^2}8
\]
for $0\leq\varepsilon\leq 2$.
According to Ref.\ \cite[p. 37]{Ledoux}, if $A\subset B^n$ is any measurable subset with $\varepsilon$-neighborhood
$A_\varepsilon$, we have
\[
   \mu_{B^n}(A_\varepsilon)\geq 1 - \frac 1 {\mu_{B^n}(A)}\cdot e^{-n\varepsilon^2/4}.
\]

Now let $L={\rm diag}(a_1,\ldots,a_n)$ be the linear map which maps the ball $B^n$ onto the ellipsoid $N$.
Since $L$ is linear, it preserves the geometric measure; that is, for every measurable subset $A\subset B^n$, we have
$\mu_N(L(A))=\mu_{B^n}(A)$. Now let $B\subset N$ be any measurable subset. We claim that
\[
   L\left[ \left(L^{-1} B\right)_{\frac\varepsilon a}\right]\subset B_\varepsilon.
\]
To prove this, let 
\begin{equation*}
	y\in L\left[ \left(L^{-1} B\right)_{\frac\varepsilon a}\right], 
\end{equation*}	
and let $x:=L^{-1}y$.
It follows that there is some $x'\in L^{-1}(B)$ such that $\|x-x'\|\leq\frac\varepsilon a$. Let $y':=Lx'\in B$,
then $\|y-y'\|=\|L(x-x')\|\leq \|L\|\cdot \|x-x'\| \leq\varepsilon$ since $\|L\|=a$, so $y\in B_\varepsilon$. Thus
\begin{eqnarray*}
   \mu_N(B_\varepsilon)&\geq& \mu_N\left(L\left[ \left(L^{-1} B\right)_{\frac\varepsilon a}\right]\right)
   =\mu_{B^n}\left(\left(L^{-1}B\right)_{\frac\varepsilon a}\right)\\
   &\geq& 1-\frac 1 {\mu_{B^n}(L^{-1}B)}\cdot e^{-n \left(\frac \varepsilon a\right)^2/4}\\
   &=&1-\frac 1 {\mu_N(B)} \cdot e^{-n\varepsilon^2/(4 a^2)}.
\end{eqnarray*}
Let now $f:N\to\R$ be any $1$-Lipschitz function, let $m_f$ be the median of $f$ on $N$, and let $A:=\{x\in N\,\,|\,\, f(x)\leq m_f\}$
such that $\mu_N(A)=\frac 1 2$. It follows that
\begin{eqnarray*}
\mu_N\{x\in N\,\,|\,\,f(x)>m_f+\varepsilon\}&\leq& \mu_N\{x\in N\,\,|\,\, x\not\in A_\varepsilon\}\\
&=& 1-\mu_N(A_\varepsilon)\\
&\leq&\frac 1 {\mu_N(A)}\cdot e^{-n\varepsilon^2/(4 a^2)}
\end{eqnarray*}
Repeating the calculation for $f(x)-m_f-\varepsilon$ and applying the union bound yields
\begin{equation}
   \mu_N\{|f(x)-m_f|>\varepsilon\}\leq 4\cdot e^{-n\varepsilon^2/(4 a^2)}=:\kappa.
   \label{EqEllConAlternative}
\end{equation}
Due to the characterization of the observable diameter as given in~(\ref{EqCharacterizationObsDiam}), we obtain
$\ObsDiam(N,\kappa)\leq 2\varepsilon$, and the claim follows from
expressing $\varepsilon$ in terms of $\kappa$.
\qed

As a last technical lemma, we need a comparison between the mean and the median on the $(n-1)$-dimensional sphere.
We suspect that the statement is well-known, but we have been unable to locate a proof in the literature.

\begin{lemma}[Mean vs. median on the sphere]
\label{LemMeanMedianSphere}
Let $f:S^{n-1}\to\R$ be any function that is $\lambda$-Lipschitz with respect to the Euclidean distance
measure, inherited from the surrounding space $\R^n$. Moreover, let $\mathbb{E}_{S^{n-1}}f$ denote the expectation
value, and $m_{S^{n-1}}f$ denote the median of $f$ on $S^{n-1}$. Then,
\[
   \left| \mathbb{E}_{S^{n-1}}f-m_{S^{n-1}} f\right| \leq \frac{\lambda\pi}{2\sqrt{n-2}}.
\]
\end{lemma}
\proof
Without loss of generality, we may assume that $\lambda=1$.
We use L\'evy's Lemma on the sphere~\cite{MilmanSchechtman}. Normally, it is formulated for the geodesic distance $\rho$
instead of the Euclidean distance $d$; but since $d\leq\rho$, it follows that $f$ must be $1$-Lipschitz
also for the geodesic distance. Since two arbitrary points on the sphere always have geodesic distance less than
or equal to $\pi$, it is clear that
$|f(x)-m_{S^{n-1}}f|\leq \pi$ for all $x\in S^{n-1}$. Let $\varepsilon>0$. L\'evy's Lemma states that
\[
   \mu_{S^{n-1}}\left\{|f(x)-m_{S^{n-1}} f|>\varepsilon\right\}\leq \sqrt{\frac\pi 2}
   \cdot e^{-\varepsilon^2(n-2)/2}.
\]
Abbreviating $m:=m_{S^{n-1}}f$ and $\mu:=\mu_{S^{n-1}}$, we get
\begin{eqnarray*}
\left| \mathbb{E}_{S^{n-1}}f-m\right| &=& |\mathbb{E}_{S^{n-1}}(f-m)|\leq \mathbb{E}_{S^{n-1}} |f-m| \\
&=&-\int_0^\infty \varepsilon \partial_\varepsilon \mu_{S^{n-1}} \{ |f-m|>\varepsilon\}\,d\varepsilon \\
&=&\int_0^\infty \mu_{S^{n-1}} \{ |f-m|>\varepsilon\}\,d\varepsilon\\
&\leq& \sqrt{\frac \pi 2} \int_0^\infty e^{-\varepsilon^2(n-2)/2} \, d\varepsilon = \frac\pi{2\sqrt{n-2}}.
\end{eqnarray*}
This proves the claim.
\qed

Now we are ready to prove concentration of measure for the manifold that arises from intersecting a sphere
with an ellipsoid in Euclidean space. 

\begin{theorem}[Concentration of measure for $M_E$, $\R$-version]
\label{TheConcentrationR}
Let $n\geq 3$, and
$\{E_k\}_{k=1}^n$ any set of positive energy levels
with arithmetic mean $E_A:=\frac 1 n \sum_{k=1}^n E_k$, harmonic mean $E_H:=\left(\frac 1 n \sum_{k=1}^n \frac 1 {E_k}\right)^{-1}$,
maximum $E_{\text{max}}:=\max_k E_k$, minimum $E_{\text{min}}:=\min_k E_k$, and $E_Q^{-2}:=\frac 1 n \sum_{k=1}^n \frac 1 {E_k^2}$.
Suppose that $E$ is any energy value which satisfies
\begin{itemize}
\item $E=\left(1+\frac 2 n\right) \left(1+\frac\varepsilon{\sqrt{n}}\right) E_H$ for some $\varepsilon>0$,
\item $E\leq E_A - \frac{\pi(E_{max}-E_{min})}{\sqrt{n-2}}$.
\end{itemize}
Then, for every $\lambda$-Lipschitz function $f:M_E\to \R$ with median $\bar f$, we have
for the normalized geometric measure $\mu\equiv \mu_{M_E}$
\begin{eqnarray*}
   \mu\left\{|f-\bar f|>t\right\}\leq
   \frac{\beta E_{\text{max}}^2 n^{\frac 3 2}}{E^2\left(1-\frac{2 E^2}{\varepsilon^2 E_Q^2}\right)}
   e^{-n\left[\frac{3 E_{\text{min}}}{64 E}\left(\frac t \lambda - \frac 1 {2n}\right)^2\right]+\varepsilon\sqrt{n}}
\end{eqnarray*}
whenever the denominator on the right-hand side is positive.
The constant $\beta>0$ can be chosen as $\beta=\frac{2048}\pi \sqrt{e}< 1075$.

Moreover, the median $\bar f$ of $f$ on $M_E$ can be estimated as follows. Let $N$ be the full ellipsoid of
points $x\in\R^n$ with $\sum_{k=1}^n E_k x_k^2\leq E\left(1+\frac 1 n\right)$,
and let $\lambda_N$ be the Lipschitz constant of $f$ in $N$. Then
\[
   \left| \bar f -\mathbb{E}_N f\right| \leq \lambda_N\left( \frac 3 {4n} + b\left({\frac E {E_{\text{min}}}\cdot \newmathcal{O}\left(n^{-\frac 1 2}\right)}\right)^\frac12\right),
\]
where the constant $b>0$ can be chosen as $b=8\left(1+\sqrt{\frac 2 3}\right)<15$, and
\[
   \newmathcal{O}\left(n^{-\frac 1 2}\right)=\frac\varepsilon{\sqrt{n}} + \frac 1 n \ln\frac{2\beta n^{\frac 3 2} E_{\text{max}}^2}
   {E^2\left(1-\frac{2 E^2}{\varepsilon^2 \bar E^2}\right)}.
\]
\end{theorem}
\proof
Let $\delta:=\frac\varepsilon{\sqrt{n}}$.
We may suppose that not all energy levels are equal, i.e., there exist $k$ and $l$ such that $E_k\neq E_l$; otherwise,
there is nothing to prove. Define the function $E:\R^n\to\R$ by $E(x):=\sum_{k=1}^n E_k x_k^2$, and
$\tilde E:\R^n\to\R$ by $\tilde E(x):=E_{\text{min}}+\sum_{k=1}^n (E_k-E_{\text{min}}) x_k^2$, then $\tilde E\upharpoonright_{S^{n-1}}=E\upharpoonright_{S^{n-1}}$.
Moreover, $\|\nabla \tilde E(x)\|^2\leq 4 (E_{\text{max}}-E_{\text{min}})^2$ for all $\|x\|\leq 1$, which proves
that the Lipschitz constant of $E\upharpoonright_{S^{n-1}}$ with respect to the Euclidean distance in $\R^n$ is
upper-bounded by $2(E_{\text{max}}-E_{\text{min}})$. Since $E_A=\mathbb{E}_{S^{n-1}} E(x)$,
the third condition together with Lemma~\ref{LemMeanMedianSphere}
ensures that $E$ is less than or equal to the median of $E(x)$ on the sphere.

For every $E'$, define the energy ellipsoid $N_{E'}$ via
$N_{E'}:=\{x\in\R^n\,\,|\,\, E(x)\leq E'\}$. Suppose that $x\in U_\varepsilon(M_E)$, so there is some $y\in M_E$ with
$\|x-y\|<\varepsilon$, hence
\begin{eqnarray*}
   |E(x)-E(y)|&\leq& \max_{z\in U_\varepsilon(M_E)} \|\nabla E(z)\|\cdot\|x-y\|\\
   &\leq& 2 E_{\text{max}}(1+\varepsilon)\varepsilon,
\end{eqnarray*}
It follows that $E(x)\leq E+4 E_{\text{max}}\varepsilon$ if $0<\varepsilon\leq 1$. Consequently,
$U_\varepsilon(M_E)\subset N_{E+4\varepsilon E_{\text{max}}}$,
and $N:=N_{E+4\varepsilon E_{\text{max}}}$ will be the surrounding body of $M_E$ that we use when applying Lemma~\ref{LemConSurroundingBody}.
We arbitrarily fix the value 
\begin{equation*}
	\varepsilon:=\frac E {4 n E_{\text{max}}} 
\end{equation*}
which turns out to be an almost optimal choice
(clearly, $0<\varepsilon\leq 1$).
Hence $N=N_{E\left(1+\frac 1 n\right)}$ is a full ellipsoid with equatorial radii 
\begin{equation*}
	a_k=\left({\frac E {E_k}\left(1+\frac 1 n\right)}\right)^\frac12.
\end{equation*}
Using that
$((1+2\delta)^{n/2}-1)^{-1}\geq e^{-n\delta}$ and a few more easy simplifications, we get by applying Lemma~\ref{LemCrucialEstimate}
\[
   \frac{\mu_{n-2}(M_E)}{\mu_n(N)}>\frac 1 2 n^{\frac 3 2} e^{-n\delta}\left(1-\frac{2 E^2}{n\delta^2 E_Q^2}\right)
   \left(1+\frac 1 n\right)^{-\frac n 2}.
\]
By Lemma~\ref{LemConEll} and $1+\frac 1 n\leq 1+\frac 1 3$, we have measure concentration in $N$:
\[
   \ObsDiam(N,\kappa)\leq 8\left({\frac 1 n\cdot \frac E {3 E_{\text{min}}} \ln\frac 4 \kappa}\right)^\frac12.
\]
Then Lemma~\ref{LemConSurroundingBody} yields measure concentration in $M_E$ -- applying
that lemma, using the previous inequalities together
with the fact that $E\leq E_{\text{max}}$ and that $\left(1+\frac 1 n\right)^{-\frac n 2}$ is decreasing and hence lower-bounded by
$\lim_{n\to\infty} \left(\left(1+\frac 1 n\right)^n\right)^{-\frac 1 2}=e^{-\frac 1 2}$, we get
\begin{eqnarray*}
   \ObsDiam(M_E,2\kappa)\leq \frac 1 {2n}+8\sqrt{\frac E {3 E_{\text{min}}}}\times\\
   \times \left({\delta+\frac 1 n\left(
      \ln\frac{512 n^{\frac 3 2} E_{\text{max}}^2}{\pi E^2 \kappa}+\frac 1 2-\ln\left(1-\frac{2 E^2}{n \delta^2 E_Q^2}\right)
   \right)}\right)^\frac12.
\end{eqnarray*}
Then the first claimed inequality follows from the characterization of the observable diameter as given in~(\ref{EqObsDiamConsequence}).

To prove the second claim, suppose that $f:N\to\R$ is any $(\lambda_N=1)$-Lipschitz function, and define for $\xi>0$
\[
   X_\xi:=\left\{x\in M_E\,\,:\,\, |f(x)-m_{M_E} f|\leq\lambda\xi\right\}.
\]
We already know that $\mu_{M_E}(X_\xi)$ is large. Lemma~\ref{LemMeasureNeighborhood} yields
\begin{eqnarray*}
   \mu_N(U_\varepsilon(X_\xi))&=&\frac{\mu_n(U_\varepsilon(X_\xi))}{\mu_n(N)} \geq \frac{\pi \varepsilon^2}{4(n-1)}
   \frac{\mu_{n-2}(X_\xi)}{\mu_n(N)}\\
   &=&\frac{\pi\varepsilon^2}{4(n-1)} \mu_{M_E}(X_\xi)\cdot\frac{\mu_{n-2}(M_E)}{\mu_n(N)}=:P.
\end{eqnarray*}
We know that we have measure concentration in $N$; setting 
\begin{equation*}
	a:=\min_k a_k=\left({\frac{E}{E_{\text{min}}}\left(1+\frac 1 n\right)} \right)^\frac12
\end{equation*}
and
using eq.~(\ref{EqEllConAlternative}) of Lemma~\ref{LemConEll}, we get for all $C>0$
\[
   \mu_N\{x\in N\,\,:\,\, |f(x)-m_N f|>C\}\leq 4\cdot e^{-n C^2/(4a^2)},
\]
and using~\cite[Appendix V.4]{MilmanSchechtman},
\[
   \mu_N\{x\in N\,\,:\,\, |f(x)-\mathbb{E}_N f|>C\}\leq 8\cdot e^{-n C^2/(32 a^2 \lambda^2)}.
\]
Set now
\[
	C:=a \left({\frac{32} n \ln\frac{8}P}\right)^\frac12,
\] 
	then $\mu_N\{|f(x)-\mathbb{E}_N f|\leq C\}
>1-P$. Thus,
\[
   U_\varepsilon(X_\xi)\cap\left\{x\in N\,\,:\,\, |f(x)-\mathbb{E}_N f|\leq C\right\}\neq\emptyset,
\]
and if $x$ is any element of that intersection, it holds $|f(x)-\mathbb{E}_N f|\leq C$ and $|f(x)-m_{M_E}f|\leq \xi+\varepsilon$, such
that
\begin{equation}
   \left| m_{M_E}f-\mathbb{E}_N f\right| \leq a \left({\frac{32} n \ln\frac{8} P}\right)^\frac12 +\xi+\varepsilon.
   \label{EqDiffEM}
\end{equation}
Now we specialize $\xi$ by setting
\begin{equation*}
   \xi:=\frac 1 {2n}+8\left({\frac E {E_{\text{min}}}
   \left(\delta+\frac\alpha n\right)
   }\right)^\frac12,
\end{equation*}
where
\[
   \alpha:=\frac 1 2-\ln\left(1-\frac{2 E^2}{n\delta^2 E_Q^2}\right)+\ln
   \frac{4096 n^{\frac 3 2}E_{\text{max}}^2}{\pi E^2}.
\]
This $\xi$ is chosen such that $\mu_{M_E}(X_\xi)\geq\frac 1 2$. The assertion of the theorem is then proved
by substituting all the previously established inequalities into~(\ref{EqDiffEM}).
\qed

Consider now the assumptions given in Theorem~\ref{MainTheorem1}, but
denote the complex dimension by $\tilde n$. Define $\tilde\varepsilon:=\frac\varepsilon{\sqrt{2}}$,
double all energy eigenvalues, and substitute this into Theorem~\ref{TheConcentrationR}. After dropping all
tildes, this substitution yields the statements of Theorem~\ref{MainTheorem1} and~\ref{MainTheorem2}.
The proof of Theorem~\ref{MainTheorem3} can now be given as follows.\\

\textit{Proof of Theorem~\ref{MainTheorem3}.} Every $|\psi\rangle\in A\otimes B$ can be written
\begin{equation*}
	|\psi\rangle=\sum_{j,k} \psi_{jk} |j\rangle\otimes |k\rangle,
\end{equation*}	
where $H_A|j\rangle=E_j^A |j\rangle$ and $H_B|k\rangle=E_k^B |k\rangle$. We embed all vectors in
real space $\R^{2n}$ by introducing coordinates $x_{jk}$ and $y_{jk}$ such that
\begin{equation*}
	\psi_{jk}=:x_{jk}+i y_{jk}. 
\end{equation*}
First we apply Theorem~\ref{MainTheorem2} to estimate the matrix
elements of $\psi^A:=\Tr_B|\psi\rangle\langle\psi|$. Embedding the ellipsoid $N$ from Theorem~\ref{MainTheorem2}
into $\R^{2n}$, we get a real ellipsoid with equatorial radii $({{E'\left(1+\frac 1 {2n}\right)}/{E'_{jk}}})^\frac 1 2$, where
each equatorial radius appears twice, namely for the coordinate axes $x_{jk}$ and $y_{jk}$.
Let $v,w\in\{x,y\}$, then a transformation to spherical coordinates yields
\begin{eqnarray*}
   \mathbb{E}_N( v_{ab} w_{cd})&=&\frac 1 {\mu_{2n}(N)} \int_N v_{ab} w_{cd} \, dz \\
   &=& \frac {E'\left(1+\frac 1 {2n}\right)}{\mu_{2n}(B^{2n}) ({E'_{ab} E'_{cd}})^\frac 1 2}
   \int_{B^{2n}} v_{ab} w_{cd} \, dz \\
   &=& \left\{
      \begin{array}{cl}
         0 & \mbox{if }(a,b)\neq (c,d) \mbox{ or }v\neq w \\
         \frac{E'\left(1+\frac 1 {2n}\right)}{E'_{ab}(2n+2)} & \mbox{otherwise},
      \end{array}
   \right.
\end{eqnarray*}
where we have used eq.~(\ref{EqIntx2bn}) and the equation $\int_{B^n} x_1 x_2 \, dx=0$. Hence
\begin{eqnarray*}
   \mathbb{E}_N \psi_{pk} \bar\psi_{qk} &=& \mathbb{E}_N \left[
      (x_{pk} x_{qk} + y_{pk} y_{qk}) + i (y_{pk} x_{qk} - x_{pk} y_{qk})
   \right]\\
   &=& \frac{\delta_{p,q} E'\left( 1+ \frac 1 {2n}\right)}{E'_{pk}(n+1)}.
\end{eqnarray*}
Since $\langle p|\psi^A|q\rangle=\sum_{k=1}^{|B|} \psi_{pk}\bar\psi_{qk}$, this yields
\[
   \mathbb{E}_N \langle p|\psi^A |q\rangle = \frac{\delta_{p,q} \left(1+\frac 1 {2n}\right)}{n+1}\sum_{k=1}^{|B|}
   \frac{E'}{E'_{pk}}.
\]
To bound the Lipschitz constants, we compute gradients: the result for the real part is
\begin{eqnarray*}
   \left\| \nabla \Re \langle p |\psi^A |q\rangle\right\|^2 &=& \left\{
      \begin{array}{c}
         \sum_{j=1}^{|B|} \left(x_{qj}^2 + x_{pj}^2 + y_{qj}^2 + y_{pj}^2\right) \\
         4\sum_{j=1}^{|B|} \left( x_{pj}^2 + y_{pj}^2\right)
      \end{array}
   \right. \\
   &\leq&\left\{
      \begin{array}{cl}
         \|\psi\|^2 & \mbox{if } p\neq q \\
         4\|\psi\|^2 & \mbox{if }p=q,
      \end{array}
   \right.
\end{eqnarray*}
and for the imaginary part, we get
\begin{eqnarray*}
   \left\| \nabla \Im \langle p |\psi^A |q\rangle\right\|^2 &=& \left\{
      \begin{array}{c}
         \sum_{j=1}^{|B|} \left(y_{pj}^2+y_{qj}^2+x_{qj}^2+x_{pj}^2\right) \\
         0
      \end{array}
   \right. \\
   &\leq&\left\{
      \begin{array}{cl}
         \|\psi\|^2 & \mbox{if } p\neq q \\
         0 & \mbox{if }p=q.
      \end{array}
   \right.
\end{eqnarray*}
Consider the functions $r_{pq}(\psi):=\Re \langle p|\psi^A|q\rangle$ and
$i_{pq}(\psi):=\Im\langle p|\psi^A |q\rangle$. All corresponding Lipschitz constants
$\lambda_N$ in the ellipsoid $N$ satisfy 
\[
\lambda_N\leq 2 \left({\frac{E'}{E'_{\text{min}}}\left(1+\frac 1 n\right)}\right)^\frac12,
\]
since this square root denotes the largest equatorial radius, which is an upper bound to $\|\psi\|$.
It follows from Theorem~\ref{MainTheorem2} that the median of both functions satisfies
\[
   \left.
   \begin{array}{c}
      |\bar r_{pq}-\Re (\rho_c)_{p,q} | \\
      |\bar i_{pq}-\Im (\rho_c)_{p,q} |
   \end{array}\right\}
   \leq 
   2 \underbrace{\left({\frac{E'}{E'_{\text{min}}}\left(1+\frac 1 n\right)}\right)^\frac12 \left(\frac 3 {8n} + 15 o_n \right)}_{=:\delta},
\]
where 
\begin{equation*}
	o_n=\left({\frac{E'}{E'_{\text{min}}}\left(\frac\varepsilon{\sqrt{n}} + \frac{\ln(2 a n^{\frac 3 2})}{2n}\right)}\right)^\frac12.
\end{equation*}	
By the triangle inequality, we have
\begin{eqnarray*}
   \left| r_{pq}(\psi)-\Re(\rho_c)_{p,q}\right|>t+2\delta&\Rightarrow&
   \left| r_{pq}(\psi)-\bar r_{pq}\right|>t \\
   \left| i_{pq}(\psi)-\Im(\rho_c)_{p,q}\right|>t+2\delta&\Rightarrow&
   \left| i_{pq}(\psi)-\bar i_{pq}\right|>t
\end{eqnarray*}
for every $t\geq 0$. Since the Lipschitz constants of
$r_{pq}$ and $i_{pq}$ in the sphere (and thus in $M_E$) satisfy $\lambda\leq 2$,
it follows from Theorem~\ref{MainTheorem1} that
\[
   {\rm Prob}\left\{ \left| r_{pq}(\psi)-\Re (\rho_c)_{p,q}\right| > 2t+2\delta\right\}\leq\eta
\]
and similarly for $i_{pq}$, where we used the abbreviation
\begin{equation*}
\eta:=a\cdot n^{\frac 3 2} \exp\left( - c n \left(t-\frac 1 {4n}\right)^2
+2\varepsilon \sqrt{n}\right). 
\end{equation*}
Combining the two inequalities for the real and the imaginary part, we get
\[
   {\rm Prob}\left\{ \left|\langle p|\psi^A |q\rangle -(\rho_c)_{p,q}\right|>\sqrt{2}(2t+2\delta)\right\}\leq 2\eta.
\]
By the union bound, the probability that there exist indices $p$ and $q$ such that
$\left|\langle p|\psi^A |q\rangle -(\rho_c)_{p,q}\right|>\sqrt{2}(2t+2\delta)$ is upper-bounded
by $|A|(|A|+1)\eta$. If this is not the case, i.e., if no such indices exist, then
\[
   \left\| \psi^A-\rho_c\right\|_2^2 = \sum_{p,q} \left(\langle p|\psi^A|q\rangle-(\rho_c)_{p,q}\right)^2
   \leq 2|A|^2(2t+2\delta)^2.
\]
Thus,
\[
   {\rm Prob}\left\{ \left\| \psi^A-\rho_c\right\|_2^2\leq 8|A|^2(t+\delta)^2\right\}\geq
   1-|A|(|A|+1)\eta.
\]
This proves the claim.
\qed

As stated in the introduction, we now give a proof that it is always possible
to shift the energy offset such that the energy $E$ in question becomes (close to) the
harmonic mean energy.
\begin{lemma}[Harmonic mean and energy shifts]
\label{LemHarmonicMean}
Suppose we are given energy levels $\{E_k\}_{k=1}^n$ and an energy $E$ between
the smallest and the arithmetic mean energy $E_A$, that is
\[
   \min_k E_k \leq E < E_A:=
   \frac 1 n \sum_{k=1}^n E_k.
\]
Then there exists $\Delta E\in\R$ such that the harmonic mean of the energies $\{E_k+\Delta E\}_{k=1}^n$
equals $E+\Delta E$, and all energies are non--negative: $E_k+\Delta E\geq 0$. Moreover, $\Delta E$ is unique
unless all energies are equal.
\end{lemma}
\proof
Denote the harmonic mean of $n$ energy values $E_1,\ldots,E_n$ by $E_H\{E_k\}_{k=1}^n:=\left(\frac 1 n\sum_{k=1}^n \frac 1 {E_k}\right)^{-1}$;
similarly, we use $E_A\{E_k\}_{k=1}^n:=\frac 1 n\sum_{k=1}^n E_k$ to emphasize the dependence of the arithmetic mean $E_A$
on the energy values.
Let $E(\Delta E):=E_H\{E_k+\Delta E\}_{k=1}^n-\Delta E$ which defines a continuous function.
We may assume without restriction that $E_1=\min_i E_i$ and $E_n=\max_i E_i$. If $\Delta E=-E_1$, then
\[
   E(\Delta E)=E_H\{E_k-E_1\}_{k=1}^n+E_1=E_H\{0,\ldots\}+E_1=E_1.
\]
It remains to show that $\lim_{\Delta E\to\infty} E(\Delta E)=E_A$, then by continuity
there must be some $\Delta E \geq -E_1$ such that $E(\Delta E)=E$. The limit identity we would like to show is equivalent to
\[
   \lim_{\Delta E\to\infty}\left(E_H\{E_k+\Delta E\}_{k=1}^n-E_A\{E_k+\Delta E\}_{k=1}^n\right)=0.
\]
We apply an inequality given by Furuta~\cite{Furuta}:
\begin{eqnarray*}
   E_A\{E_k+\Delta E\}_{k=1}^n\geq E_H\{E_k+\Delta E\}_{k=1}^n\\
   \geq
   \underbrace{\frac{4(E_1+\Delta E)(E_n+\Delta E)}{(E_1+E_n+2\Delta E)^2}}_{\to 1\mbox{ for }\Delta E\to\infty}
   E_A\{E_k+\Delta E\}_{k=1}^n.
\end{eqnarray*}
This proves existence of some $\Delta E$ such that $E(\Delta E)=E$. In order to see uniqueness, note that the Cauchy-Schwarz
inequality for two vectors $v=\left(\frac 1 n,\ldots,\frac 1 n\right)^T$ and $a=(a_1,\ldots,a_n)^T$, i.e.
$|\langle v,a\rangle|^2\leq \langle v,v\rangle\langle a,a\rangle$ implies that $\left(\frac 1 n \sum_{k=1}^n a_k\right)^2
\leq \frac 1 n \sum_{k=1}^n a_k^2$, and we have equality if and only if $a_1=a_2=\ldots=a_n$.
The derivative of the function $E$ turns out to be
\[
   E'(x)=\left(\frac 1 n \sum_{k=1}^n \frac 1 {(E_k+x)^2}\right)/\left(\frac 1 n \sum_{k=1}^n \frac 1 {E_k+x}\right)^2 -1
\]
which must thus be strictly positive unless all energies are equal.
\qed

\section{Approximate sampling of the manifold}
\label{SecGaussian}
There is a well-known method~\cite{Muller,Marsaglia} to pick random points from the surface of a hypersphere $S^{n-1}\subset \R^n$:
Generate $X_1,X_2,\ldots,X_n$ random real numbers, distributed independently identically according
to the normal distribution with density proportional to $\exp(-n x^2/2)$. Then, normalize the resulting vector:
For $r:=({X_1^2+X_2^2+\ldots + X_n^2})^\frac12$, the point
\[
   \left(X_1,X_2,\ldots,X_n\right)^T/r
\]
is uniformly distributed on the unit hypersphere.

If the uniform distribution on the hypersphere shall be sampled only \emph{approximately}, then
the normalization is not necessary: we have the norm expectation value $\mathbb{E}\|X\|^2=1$ and
the variance ${\rm Var}\|X\|^2=\frac 2 n$, such that the distribution of the vectors $X$ themselves
closely resembles the uniform distribution on the sphere in high dimensions $n$. This way,
expectation values of functions $f:\R^n\to\R$ with respect to the uniform distribution on the sphere
can be estimated numerically to good accuracy (assuming that $f$ is slowly varying and not growing too fast at infinity).
This has the quantum interpretation (if $n$ is even) of drawing random pure states in $\C^{n/2}$.

It turns out that a simple modification of this algorithm yields \emph{approximate sampling
of the energy manifold $M_E$}, or rather of its measure $\mu_{M_E}={\rm Prob}$ that we use in this paper.
We describe the algorithm below. In contrast to the rest of this paper,
we do not give explicit error bounds in this case, because the necessary calculations are straightforward but
very lengthy, and the resulting error bounds depend sensitively on the assumptions on the
regularity of the functions $f$ that are considered. However, we discuss a rough estimate of the error at the
end of this section.

\begin{algorithm}[Approximate sampling of $M_E$]
\label{algorithm}
Suppose we are given an observable $H=H^\dagger$ on $\C^n$ with eigenvalues $\{E_k\}_{k=1}^n$ and an energy value $E$
such that Theorem~\ref{MainTheorem1} applies and proves sufficient concentration of measure.
Then, the uniform (Hausdorff) measure on the manifold of quantum states $|\psi\rangle$
with $\langle\psi|H|\psi\rangle=E$ can be numerically sampled in the following way:
\begin{itemize}
\item[1.] Find an energy shift $s\in\R$ such that $H':=H+s\idn\geq 0$, and
such that the harmonic mean of the new energy levels $E'_k:=E_k+s$ equals $E':=E+s$, i.e.,
\[
   E'_H:=\left(\frac 1 n\sum_{k=1}^n \frac 1 {E'_k}\right)^{-1}=E'.
\]
\item[2.] Generate all real and imaginary parts $\Re\psi_k$ and $\Im\psi_k$ of $|\psi\rangle$ (in the eigenbasis
of $H$) independently according to the Gaussian distribution proportional to
\[
   \exp\left(-n \frac {E'_k}{E'} x^2\right).
\]
\end{itemize}
\end{algorithm}
If $H=\idn$ and $E=1$ (which means that we have a void constraint), this algorithm reduces to the
well-known sphere point picking algorithm as a special case (note that the real dimension is $2n$, which
cancels a factor $\frac 1 2$ in the exponent). If $H$ is not proportional to the identity, then
the entries of the random vector $|\psi\rangle$ are independently, but not identically distributed.
Note that a similar ``Gaussian approximation'' has been used in Ref.\ \cite{FreschMoro} right from the start
in the analysis of the mean energy ensemble (without error bounds).

Without using the results in this paper, direct calculation shows that the distribution generated by the algorithm above satisfies
\[
	\mathbb{E}\|\psi\|^2=\frac {E'}{E'_H}=1 
\]	
(explaining the choice of the energy shift) and
$\mathbb{E}\langle\psi|H'|\psi\rangle=E'$. The
corresponding variances are ${\rm Var}\langle\psi|H'|\psi\rangle=\frac{{E'}^2}n$ and
\begin{equation*}
	{\rm Var}\|\psi\|^2=\frac 1 {n^2}\sum_{k=1}^n \left(\frac{E'}{E'_k}\right)^2; 
\end{equation*}
this expression
is also present in Theorem~\ref{MainTheorem1}, where it is called ${E'}^2/(n {E'}_Q^2)$
and assumed to be small (the factor $n$ is absorbed into $\varepsilon$ there).

Thus, the algorithm above produces points close to the energy manifold $M_E$ with high probability.
But does it approximate the \emph{uniform} distribution on $M_E$? Since physics mainly involves
computing expectation values of observables, we are interested in a weak form of approximation
where we say that two measure $\mu$ and $\nu$ on $\C^n\sim \R^{2n}$ (or on submanifolds) are close, i.e., 
$\mu\approx\nu$, if $\mathbb{E}_\mu f
\approx \mathbb{E}_\nu f$ for all real functions $f:\C^n\to\R$ that satisfy certain regularity conditions
(such as Lipschitz continuity and polynomial growth at infinity).

For example, the uniform measure on the sphere $\mu_{S^{2n-1}}$ and in the ball $\mu_{B^{2n}}$ are close if
$n$ is large: Since
\[
   \mu_{B^{2n}}\left\{|\psi\rangle\,\,|\,\,\|\psi\|<1-\varepsilon\right\}=(1-\varepsilon)^{2n}\leq \exp(-2n\varepsilon),
\]
most of the points in the ball are close to the surface. As a consequence, a simple calculation shows that expectation values of
$\lambda$-Lipschitz functions $f:B^{2n}\to\R$ satisfy
\[
   \left|\mathbb{E}_{B^{2n}} f - \mathbb{E}_{S^{2n-1}} f \right| \leq \frac \lambda {2n+1}.
\]
Are the uniform measure $\mu_{M_E}$ and the resulting Gaussian measure from Algorithm~\ref{algorithm} close in this sense?
The answer is yes, and the results
in this paper give a simple geometric explanation for this fact, which is schematically depicted in Figure~\ref{fig:sampling}:
\begin{figure}
\includegraphics[width=0.35\textwidth]{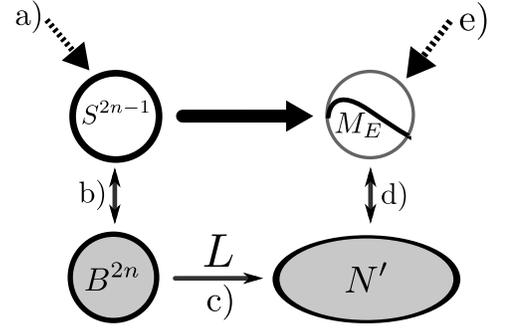}
\caption{Geometric caricature why Algorithm~\ref{algorithm} samples the measure $\mu_{M_E}$ on the energy
manifold to good approximation. Point a) denotes the well-known sphere point picking algorithm,
and e) is for Algorithm~\ref{algorithm}. See the text below for an explanation how the one leads to the other.
}
\label{fig:sampling}
\end{figure}
\begin{itemize}
\item[a)] As explained at the beginning of this section, it is well-known how to pick random points approximately from the uniform
distribution on the sphere $S^{2n-1}$: choose real and imaginary parts randomly, distributed independently identically
according to a Gaussian distribution proportional to $\exp(-n x^2)$.
\item[b)] We have just seen that the uniform distribution in the ball $B^{2n}$ and on the sphere $S^{2n-1}$ are close
to each other. Hence the algorithm from a) also samples the uniform distribution in the ball to good approximation.
\item[c)] Let $N'$ be the full ellipsoid with equatorial radii 
\begin{equation*}
\{a_k\}_{k=1}^{n}:=\left\{({E'/E'_k})^\frac12 \right\}_{k=1}^{n}
\end{equation*}
in the directions of the eigenvectors of $H'$, i.e.,
\[
   N':=\left\{z\in\C\,\,|\,\, \langle z|H'|z\rangle\leq E'\right\}.
\]
Then the ball $B^{2n}$ and the ellipsoid $N'$ are related by a linear transformation $L:\C^n\to\C^n$ which
preserves the normalized geometric volume measure: $L:={\rm diag}(a_1,\ldots,a_n)$, then
$N'=L\, B^{(2n)}$ and $\mu_{B^{2n}}=L^*\left(\mu_{N'}\right)$.

Sampling the ball $B^{2n}$, and then applying the linear transformation $L$, is the same as
sampling the full ellipsoid $N'$. Writing $y=Lx$, the components $y_k$ of vectors after the
transformation are related to the components $x_K$ before by $y_k=({E'}/{E'_k})^\frac12 x_k$. Hence
\begin{eqnarray*}
   \exp\left(-n x_k^2\right)&=&\exp\left(-n\left(\left({\frac{E'_k}{E'}}\right)^\frac12\, y_k\right)^2\right)\\
   &=& \exp\left(-n\frac{E'_k}{E'} y_k^2\right).
\end{eqnarray*}
Thus, we have shown that Algorithm~\ref{algorithm} samples the full ellipsoid $N'$ to good approximation.
\item[d)] The full ellipsoid $N'$ is close the the full ellipsoid $N$ from Theorem~\ref{MainTheorem2}.
There, it was shown that the uniform volume measure in $N$ is close to the uniform measure $\mu_{M_E}$ on
the energy manifold $M_E$. Hence the uniform volume measure in $N'$ is close to $\mu_{M_E}$.
\item[e)] In summary, the measure produced by Algorithm~\ref{algorithm} is close
to $\mu_{M_E}$ as claimed -- assuming that the underlying Hamiltonian is in the range of applicability of
Theorem~\ref{MainTheorem1} and~\ref{MainTheorem2}.
\end{itemize}

The sampling algorithm gives a simple method for a numerical check of identities such as the form of the
reduced density matrix in Example~\ref{ExMainRed}. It is interesting to note that even though
the typical reduced density matrices are \emph{not} Gibbs states (cf.\ Theorem~\ref{MainTheorem3}),
the distribution involved in Algorithm~\ref{algorithm} involves the Boltzmann-like term
$\exp(-c E'_k)$, where $c>0$ is constant, and $E'_k$ denotes the $k$-th energy level.

How good is the approximation given by Algorithm~\ref{algorithm}? As explained above, the algorithm is
meant to approximate expectation values of functions $f:\R^n\to\R$ on the energy manifold with respect to
the measure $\mu_{M_E}$. Thus, we would like to estimate the expression $|\mathbb{E}f-\mathbb{E}_{M_E} f|$,
where $\mathbb{E}$ denotes the expectation value with respect to the Gaussian measure used in the
algorithm. As a lower bound on that error (for some $f$), recall that
\[
   {\rm Var}\|\psi\|^2=\mathbb{E}\left(\|\psi\|^2-1\right)^2=\frac{{E'}^2}{n {E'}_Q^2}.
\]
The function $f(\psi):=\left(\|\psi\|^2-1\right)^2$ is Lipschitz continuous, and the
upper bound on the Lipschitz constant $\|\nabla f(\psi)\|=4\|\psi\|(1-\|\psi\|^2)\leq 8/(3\sqrt{3})$
on the unit ball does not grow with $n$. Assume for simplicity that $E'$ and $E'_Q$ are constant in $n$
(like in Example~\ref{ExMain1} and Example~\ref{ExMainRed}). Then
\[
   |\mathbb{E} f -\mathbb{E}_{M_E} f|=\frac{{E'}^2}{{E'}_Q^2}\cdot\frac 1 n,
\]
which shows that we have to expect at least an error of the order $1/n$ even for functions $f$ and Hamiltonians $H$ that behave very regularly.

A rough upper bound on the error can be given by adding the error contributions of steps a), b), c), and d) in
Algorithm~\ref{algorithm}. It seems that the dominant contribution comes from step d) -- a corresponding error
estimate is given in Theorem~\ref{MainTheorem2}. It is roughly of the order $n^{-1/4}$, again assuming that the energies
and the Lipschitz constant are constant in $n$.

\section{Conclusions}

In this work, we have established  the notion of concentration of measure for quantum
states with a fixed expectation value. The results that we established constitute on the 
one hand a new proof tool to assess properties of quantum states with the probabilistic method.
Such a proof tool is expected to be helpful in a number of contexts, e.g., when sharpening 
counterexamples to additivity by enforcing a strong ``conspiracy'' by means of a suitable
Hilbert Schmidt constraint, adding to the portfolio of techniques available related to
the idea of a probabilistic method. 

On the other hand, in this work we are in the position to introduce concentration
of measure ideas to notions from quantum statistical mechanics,
specifically to the mean energy ensemble, and link this physically meaningful
ensemble to ideas of typicality. Obviously, a constraint of the type introduced here could as well
relate to settings where the particle number is held constant, so is expected to be applicable
to a quite wide range of physical settings. It is also the hope that methods similar to the 
ones established here also help assessing questions of typicality in the context of quantum dynamics 
and addressing key open problems in the theory of {\it relaxation}
\cite{Kollath,Rigol,Exact,Speed} of non-equilibrium complex quantum systems.

\section{Acknowledgments}
We would like to thank C.\ Gogolin and R.\ Seiler for discussions. 
This work has been supported by the EU (COMPAS, CORNER, QESSENCE, MINOS) and the EURYI.

\end{document}